\documentclass{aa}  

\usepackage{graphicx}
\usepackage{xcolor}
\usepackage{subfigure}
\usepackage{subcaption}

\usepackage[switch]{lineno}

\usepackage{txfonts}
\definecolor{myblue}{rgb}{0.00, 0.0, 0.9}
\definecolor{myred}{rgb}{0.90, 0.0, 0.0}
\definecolor{mygreen}{rgb}{0.0, 0.7, 0.0}
\usepackage[breaklinks,  citecolor=myblue, linkcolor=myred, urlcolor=purple, colorlinks=true, debug, baseurl=' ']{hyperref}
\usepackage{orcidlink}
\usepackage{ulem}

\begin{document} 
%\linenumbers

   \title{Linear polarization study of open clusters towards the anticenter direction: Signature of the spiral arms\thanks{Full Table \ref{tab:obs}, and the tables containing the polarization measurements of the remaining four clusters are available at the CDS via anonymous ftp to \href{https://cdsarc.u-strasbg.fr/}{cdsarc.u-strasbg.fr
(130.79.128.5)} or via \href{http://cdsweb.u-strasbg.fr/cgi-bin/qcat?J/A+A/}{http://cdsweb.u-strasbg.fr/cgi-bin/qcat?J/A+A/}.}}

\author{Namita Uppal\inst{1,2,3}
          \and
          Shashikiran Ganesh\inst{2}
          \and 
          Vincent Pelgrims\inst{4}
          \and
          Santosh Joshi\inst{5}
          \and
          Mrinmoy Sarkar\inst{5}
          }

   \institute{Institute of Astrophysics, Foundation for Research and Technology-Hellas, GR-71110 Heraklion, Greece \\
   \email{namita@ia.forth.gr}
   \and
   Physical Research Laboratory, Ahmedabad, 380009, Gujarat, India
         \and
             Indian Institute of Technology, Gandhinagar, 382355, Gujarat, India
        \and
            Universit{\'e} Libre de Bruxelles, Science Faculty CP230, B-1050 Brussels, Belgium
        \and
            Aryabhatta Research Institute of Observational Sciences, Manora Peak, Nainital 263002, India\\
             }

% \abstract{}{}{}{}{} 
% 5 {} token are mandatory
  \abstract
   {}
   {Our objective is to investigate the distribution of dust and associated large-scale structures of the Galaxy using optical linear polarization measurements of various open clusters located at different distances in the Galactic anticenter direction.  
   }
   {We present R-band linear polarization observations of stars towards five open clusters: Kronberger~1, Berkeley~69, Berkeley~71, Berkeley~19, and King~8 in the anticenter direction. The polarization observations were carried out using ARIES IMaging POLarimeter mounted on the 104 cm Sampurnanand telescope of ARIES, Nainital, making it the first study to target the polarization observations towards distant clusters ($\sim$6~kpc). We combined the observed polarization data with the distance information from the \textit{Gaia} space telescope to infer the dust distribution along the line of sight. }
   {The variation in the degree of polarization and extinction with distance reveals the presence of multiple dust layers along each cluster direction. In addition, common foreground dust layers detected towards different cluster directions highlight the presence of global features such as spiral arms. Our results show that the dust clouds at 2~kpc towards Berkeley~69 and Berkeley~71 coincide with the Perseus arm, while the dust layer at $\sim$4~kpc towards distant clusters, Berkeley~19 and King~8, indicates the presence of the Outer arm.  The large-scale dust distribution obtained by combining our polarization results with the previous polarization studies of nearby open clusters suggests that the anticenter direction is characterized by low extinction, homogeneous dust distribution with somewhat uniform orientation of the plane-of-sky component of the magnetic field along the line of sight.}
   { Our study demonstrates the utility of polarization as a tool to study the large-scale dust distribution and structural features where the kinematic distance methods are inadequate to provide accurate distances to the dust clouds. The global dust distribution towards the anticenter direction shows signatures of the intervening spiral arms.}

   \keywords{(Galaxy:) open clusters and associations: individual: Kronberger~1, Berkeley~69, Berkeley~71, Berkeley~19, King~8 -- Galaxy: structure -- ISM: general -- (ISM:) dust, extinction -- Methods: observational -- Techniques:polarimetric
               }

   \maketitle
%
%-------------------------------------------------------------------
%@arxiver{Final/Fig6.pdf, Final/Fig7.pdf}
\section{Introduction}\label{sec:1}
Tracing the distribution of dust in the Galaxy is of paramount importance for mapping its structure. Numerous attempts have been made to create three-dimensional extinction maps of the Milky Way galaxy to uncover the dust distribution \citep[e.g.,][]{Lallement2019, Green2019, Lallement2022, vergely2022, edenhofer2023}. However, these maps rely on models and assumptions, possibly introducing inherent biases, such as assumptions regarding the intrinsic Spectral Energy Distributions (SEDs) of the stars utilized for extinction determination. Another notable property demonstrated by dust grains is starlight polarization. An unpolarized light beam, when passed through the asymmetrical but aligned dust grains, suffers dichroic extinction. 
In other words, when unpolarized light from the stars interacts with partially aligned,  optically anisotropic (e.g., elongated) dust grains, a part of the light is absorbed by these grains. The electric field vector parallel to the longer axis of the grain gets preferentially absorbed due to its larger cross-sectional area. Consequently, the scattered light becomes partially plane polarized along the short axis of the grain \citep{Hall1949, Hiltner1949}. Moreover, the dust grains are aligned with their shorter axis parallel to the ambient magnetic field \citep{Davis1949} direction. Hence, the polarization angles indicate the plane-of-sky orientation of the magnetic field.
In a broader context, the polarization of starlight caused by interstellar dust is regarded as a valuable tool to determine various properties of the interstellar medium (ISM), such as the geometry of the magnetic field \citep[e.g.,][]{mathewson1970polarization, goodman1990optical, GPIPS} and the grain properties like size, shape, chemical composition, and grain alignment efficiency \citep[e.g.,][]{kim1994size, efficiency, NewSerkowski}. Furthermore, combining polarization data with accurate distance information provides observational constraints on the number of dust clouds and their distances along the line of sight \citep[e.g.,][]{Eswaraiah2012, panopoulou2019, Doi2021, Uppal2022, Bijas2022, Doi2023, Pelgrims2023, Pelgrims2024}, thereby enabling a more comprehensive understanding of its distribution within the Galaxy.

Open clusters are essential constituents of the Galactic disk, and their distribution has been studied as a means of tracing the spiral arm structure of the Galaxy \citep[e.g.,][]{Ginard2021, spiral2}. These clusters are ideal for polarization observations as all of their member stars possess common properties such as distance, age, and proper motion. Moreover, a bunch of stars present at similar distances and locations in the sky will provide statistically more robust results than randomly selected stars.
The polarization data from various clusters can be utilized to examine the dust distribution in the Galactic disk, particularly in the spiral arms. Polarization studies of open clusters have been used to identify the number of foreground dust layers, constrain the distance of foreground clouds \citep[e.g.,][]{Eswaraiah2012, Alessi1, NGC1817, Uppal2022, Sadhana2022}, cluster membership \citep{membership}, and to gain insight into the properties of the dust \citep[e.g.,][and references therein]{NGC654, Eswaraiah2011,  Eswaraiah2012, NGC1502, Alessi1, NGC1817, Sadhana2022, Bijas2024}. The linear polarization study of open clusters combined with the distance information from the \textit{Gaia} early data release 3 \citep[EDR3,][]{EDR3} can assist in tracing the dust distribution.

Motivated by the idea of tracing the spiral arms from the dust distribution, we choose to study several clusters in a similar line of sight but present at different distances towards the anticenter direction. In contrast to the Galactic center direction, the anticenter direction is less crowded and has less dust extinction \citep{GaiaAnticenter}, providing a unique opportunity to perform polarization observations of open clusters at distant locations of the disk.  
Notably, the kinematic distances derived from the radial velocity measurements of dust clouds at longer wavelengths in anticenter directions do not provide reliable results \citep[e.g.,][]{Wenger2018, Hunter2024} which is attributed to the very small radial velocity component in this particular direction. Consequently, employing the polarization of background starlight in combination with distance emerges as the most effective method for determining the distance to foreground dust layers in this direction. 

In the literature, only five clusters have been studied in terms of their polarization measurements in the anticenter direction. These include NGC~2281 \citep{Eswaraiah2011}, NGC~1960 \citep{Eswaraiah2011}, Stock~8 \citep{Eswaraiah2011}, NGC~1931 \citep{Pandey2013}, and NGC~1893 \citep{Eswaraiah2011, Bijas2022}. However, it is noteworthy that most of these clusters are located at distances less than 2.5 kpc. To investigate the dust distribution in the distant regions, we targeted five Galactic open clusters, Kronberger~1, Berkeley~69, Berkeley~71, Berkeley~19, and King~8, towards a similar line of sight and distributed according to distance. These clusters span a distance range of $\sim 2-6$ kpc (see the sixth column of Table~\ref{tab:1}).

The paper is organized as follows. Section~\ref{sec:two} delves into the observations of the five clusters, the data reduction procedure, and details about the archival data used to complement the observations.  Section~\ref{sec:4} describes the polarization results based on the sky map and the distribution of stars in the degree of polarization and polarization angle ($P-\theta$) plane. The discussion of the dust distribution at local and global scales, utilizing the polarization results along with archival cluster polarimetric data, is described in Sect.~\ref{sec:5}. Finally, the conclusion of the paper is presented in Sect.~\ref{sec:6}.

\section{Data}\label{sec:two}
\subsection{Observations and data reduction}\label{sec:2}
The observations of five clusters, Kronberger~1, Berkeley~69, Berkeley~71, King~8, and Berkeley~19, were carried out on dark nights from October 19, 2022, to October 23, 2022. An observation log is provided in Table~\ref{tab:1} with clusters listed in ascending order of their distance, and the same ordering is followed in the rest of the text.  The polarization observations were conducted using the ARIES IMaging POLarimeter \cite[AIMPOL,][]{Rautela2004, Pandey2023} instrument mounted at the Cassegrain focus of 1.04~m, f/13 Sampurnanad telescope. The instrument consists of an achromatic rotating half wave-plate (HWP), which helps to modulate the intensity of the star, and a Wollaston prism, which is used as an analyzer. The Wollaston splits the incoming light into ordinary (o-ray, $I_o$) and extraordinary rays (e-ray, $I_e$), which are focused on the same detector plane but with a separation of $\sim$34 pixels.  A liquid nitrogen-cooled pylon CCD,  having dimensions 1300 $\times$ 1340 and pixel size of  $20 \mu$m $\times 20 \mu$m, is used to detect the light passing through the instrument, giving a total field of view of $\sim$8$^{\prime}$.  Different readout speeds (50, 100, 200, 500 kHz, 1, 2,  4 MHz) and gain (low, medium, and high) options are available in the CCD.  
The observations of all the clusters were performed in Johnson-Cousins R-band ($\lambda_{eff} = 0.63 \mu m$), with the readout speed of 100 kHz (noise 4 $e^{-}$) and medium gain. In order to obtain the normalized Stokes parameters $q = Q/I$ and $u = U/I$, we took images at four independent HWP orientations \citep[$0^\circ$, $22.5^\circ$, $45.0^\circ$, and $67.5^\circ$;][]{Schaefer2007}. Different exposure times, based on the cluster brightness, were used to cover the bright as well as the faint stars in the field up to 17~mag. The details of the exposure time for each cluster are presented in the fifth column of Table~\ref{tab:1}. At least six frames at each position were taken in order to increase the signal-to-noise ratio. 
We have observed an area of $6^\prime$ radius around each cluster, completely covering the core regions of the cluster  \citep[core size = $2.4^\prime$, $3^\prime$, $2.4^\prime$, $2.4^\prime$, $1.8^\prime$  for Kronberger~1, Berkeley~69, Berkeley~71, Berkeley~19 and King~8 respectively;][]{kharchenko2013}. Two to five polarized standards (see Table~\ref{tab:2}) and multiple sets of un-polarized standards (HD14069, Table~\ref{tab:unpol}) were also observed on each night along with the cluster field in the same filter to calibrate the polarization angle and to correct for the instrumental polarization, respectively. The instrumental polarization was found to be small ($\sim 0.1\%$). 

\begin{table*}
\caption{Observation log for open clusters observed towards anticenter direction from AIMPOL along with the distance to the respective cluster \citep{Hunt2023}. }\label{tab:1}
    \centering{
    \small
    \begin{tabular}{l c c c c c}
	\hline
        \hline \\[-2.ex]
        \setlength\tabcolsep{0.5pt}
	Cluster & RAJ2000 & DECJ2000 & Obs. date & exposure time\tablefootmark{*} & distance \\ 
        &   (deg) & (deg) & & (s) & (kpc)\\ \\[-2.ex]
	\hline \\[-1.5ex]
        Kronberger~1 & 82.089 & 34.777 & Oct. 23, 2022 & 60, 30, 10 & 2.12\\
        Berkeley~69 & 81.090 & 32.607 & Oct. 22, 2022 & 60, 20, 01 & 3.18\\
        Berkeley~71 & 85.233 & 32.272 & Oct. 19, 2022 & 60, 30 & 3.50\\
        King~8 & 87.324 & 33.633 & Oct. 23, 2022 & 75, 60, 30 & 4.79\\ 
        Berkeley~19 & 81.014 & 29.575 & Oct. 21, 2022 & 75, 60 & 5.25\\
        \\[-2.ex]
	\hline \\[-1.5ex]
% \multicolumn{4}{l}{\textsuperscript{*}\footnotesize{The exposure time (in seconds) set at each position of HWP.}}
	\end{tabular}
 \newline}
\tablefoot{\tablefoottext{*}{The exposure time (in seconds) set at each position of HWP. Multiple exposure times were used to cover bright as well as faint stars.}}

\end{table*}

The observed data were reduced and analyzed using self-scripted Python routines. The basic data reduction tasks include - bias subtraction, shifting, and stacking of images to increase the signal-to-noise ratio and astrometry using the "astrometry.net" service. The astrometric accuracy of all the images is better than 0.5$^{\arcsec}$.
Since the two images ($e$-ray and $o$-ray) of one source are present on the same CCD plane, it is important to separate the $o$-ray and $e$-ray images. Following the extraction of $e$-ray and $o$-ray coordinate information, an aperture photometry is performed on each set of coordinates using the photutils package of \citep{photutils} of \textit{Astropy} \citep{astropyII}. In crowded fields like open clusters, there is always a finite probability of apertures of nearby stars to overlap. %overlapping the nearby star. 
The crowding problem in AIMPOL images becomes multi-fold due to the presence of both $e$-ray and $o$-ray images \citep[see Fig.~4 of ][]{uppal2023optical}. Multiple apertures between 1 $\times$ FWHM to 3 $\times$ FWHM were used for aperture photometry to reduce the overlapping issue. The modulation factor R($\alpha$) at each HWP orientation ($\alpha$ = $0^\circ$, $22.5^\circ$, $45^\circ$, and $67.5^\circ$) is calculated using the photometric flux of both $e$-ray and $o$-ray ($I_o$ and $I_e$) of a star at each aperture using:% Eq.~\ref{eq:1}.
\begin{equation}\label{eq:1}
    R(\alpha) =  \frac{\frac{I_e}{I_o} \times F-1}{\frac{I_e}{I_o}\times F+1}\;, 
\end{equation}
where the factor $F$ is introduced because
the responses of the CCD to the two orthogonal polarization components may not be the same and are a function of position on its surface. The actual measured signal in the two images may differ by a factor $F$, given below:

\begin{equation}\label{eq:2}
    F = \left[\frac{I_o(0^\circ)}{I_e(45^\circ)} \times \frac{I_o(45^\circ)}{I_e(0^\circ)}\times \frac{I_o(22.5^\circ)}{I_e(67.5^\circ)}\times\frac{I_o(67.5^\circ)}{I_e(22.5^\circ)}\right]^{\frac{1}{4}}\;.
\end{equation}
The uncertainty in the measurement of R($\alpha$) is obtained using: % Eq.~\ref{eq:3}.
\begin{equation}\label{eq:3}
    \sigma_{R(\alpha)} =  \frac{\sqrt{I_e + I_o + 2I_b}}{I_e + I_o} \;, 
\end{equation}
where $I_b$ is the average background counts around the extraordinary and ordinary images.

For AIMPOL, the modulation factor measured at different HWP angles relates to the Stokes parameter of the starlight ($q_{obs}$, and $u_{obs}$) through:
\begin{equation}\label{eq:4}
    R(\alpha) =  q_{obs}\cos 4 \alpha + u_{obs}\sin 4 \alpha \; .
\end{equation}
We keep the Stokes parameters corresponding to the aperture leading to the minimum $\chi^2$ fit of Eq.~\ref{eq:4}. The obtained Stokes parameters of all the observed stars ($q_{obs}$ and $u_{obs}$) and their uncertainties ($\sigma_{q_{obs}}$ and $\sigma_{u_{obs}}$) were corrected for instrumental polarization in $qu$-space following Eq.~\ref{qu} and Eq.~\ref{equ}. 
\begin{equation}\label{qu}
    q = q_{obs} - q_{ins} \;, \\
    u = u_{obs} - u_{ins} \; , 
\end{equation}

\begin{equation}\label{equ}
    \sigma_q = \sqrt{\sigma_{q_{obs}}^2 + \sigma_{q_{ins}}^2} \;,\\
     \sigma_u = \sqrt{\sigma_{u_{obs}}^2 + \sigma_{u_{ins}}^2} \; , 
\end{equation}
where $q_{ins}$ and $u_{ins}$ are Stokes parameters corresponding to the instrumental polarization obtained from the measurement of unpolarized standard stars. $\sigma_{q_{ins}}$ and $\sigma_{u_{ins}}$ are their respective uncertainties.
The degree of polarization is then determined as the quadrature sum of $q$ and $u$ Stokes parameters, i.e., 
\begin{equation}\label{eq:5}
    p_{obs} =  \sqrt{q^2 + u^2} \; ,
\end{equation}
and the polarization angle, measured in IAU convention (North-to East), i.e., with zero at North and increasing towards the East, is determined as: 
\begin{equation}\label{eq:6}
  \theta = \frac{1}{2}\; {\rm{arctan2}}(u,\,q) \;.
\end{equation}
%Vincent: adding this
Here, in defining polarization angle, we use the two-argument arctangent function to account for the inherent $180^\circ$ ambiguity in polarization angle. 
The observed position angle of stars in each field was corrected to the reference position angle using observations of the polarized standard stars. The measured degree of polarization, polarization angle, and offset in angle for different standard stars observed on each night are listed in Table~\ref{tab:2} in $-90 \le \theta < +90$ range.

The uncertainties in the derived degree of polarization (Eq.~\ref{eq:5}) and polarization angle (Eq.~\ref{eq:6}) are estimated using error propagation of measured $q$ and $u$ values and their uncertainties using:% Eqs.~\ref{eq:5a} and~\ref{eq:6a}.
\begin{equation}\label{eq:5a}
 \sigma_{p} =  \sqrt{\frac{q^2\sigma_{q}^2 + u^2\sigma_{u}^2}{q^2+u^2}} \;,
\end{equation}
\begin{equation}\label{eq:6a}
 \sigma_{\theta} = \frac{1}{2(q^2+u^2)} \sqrt{q^2\sigma_{u}^2 + u^2\sigma_{q}^2} \; .
\end{equation}
It is noted that the determination of the degree of polarization from the quadrature sum of $q$ and $u$ Stokes parameters (Eq.~\ref{eq:5}) follows the Ricean distribution.
The presence of errors in $q$ and $u$, originating from various sources, contribute positively to the overall error, leading to a systematic increase in the derived degree of polarization \citep{Patat2006, Sohn2011}.  Consequently, this introduces a bias in the polarization measurements, particularly when the error in polarization becomes comparable to the degree of polarization. We corrected for the Ricean bias produced due to the low signal-to-noise ratio (S/N) in polarization measurements with the asymptotic estimator \citep{wardle1974} given by:
\begin{equation}\label{eq:7}
    P = \sqrt{p_{obs} ^2 - \sigma_{p} ^2}\;.
\end{equation}
The debiased value of the degree of polarization obtained from the above equation is used in our analysis. On the other hand, the measurements of polarization angle follow a different probability distribution than the degree of polarization \citep{Naghizadeh1993}. Given that the distributions of $q$ and $u$ exhibit similar statistical properties, the ratios between them do not influence the derived polarization angle (see Eq.~\ref{eq:6}). Therefore, we anticipate that measurements of polarization angle will be considerably more accurate than those of the degree of polarization. Consequently, we did not consider any debiasing in polarization angle. 

\begin{table*}
	\centering
	\caption{Polarization measurements and the reference values of highly polarized standard stars.}
	\label{tab:2}
        %\small
	\begin{tabular}{lcccccr} % four columns, alignment for each
 \hline
		\hline
		Observation date & Star & P $\pm$ $\sigma_P$ ($\%$) & $\theta_{obs}$ $\pm$ $\sigma_{\theta}$ ($^\circ$) & P $\pm$ $\sigma_P$ ($\%$) & $\theta$ $\pm$ $\sigma_{\theta}$ ($^\circ$) & angle offset \\
    & & \multicolumn{2}{c}{(our work)} & \multicolumn{2}{c}{\citep{HD25443}} & ($^\circ$) \\
    \hline
		\hline
            \textbf{Oct 19, 2022} & & & & & &\\
            \hline
            &Hiltner 960 & 5.53 $\pm$ 0.14 & -25.16 $\pm$ 0.90 & 5.210 $\pm$ 0.029 & 54.54 $\pm$ 0.16 & 79.70\\
            &HD236954 & 6.13 $\pm$ 0.15 & 30.78 $\pm$ 0.84  & 5.790 $\pm$ 0.099 & -68.80 $\pm$ 0.49 & 80.42\\
            &BD +64106 &  5.35 $\pm$ 0.07 & 18.18 $\pm$ 0.37 & 5.150 $\pm$ 0.098 & -83.26 $\pm$ 0.54   & 78.56\\
            &HD19820 & 4.32 $\pm$ 0.14 & 36.38 $\pm$ 0.76& 4.562 $\pm$ 0.025 & -65.54 $\pm$ 0.16 & 78.08 \\
            \hline
            \textbf{Oct 21, 2022} & & & & &  &\\
            \hline
            &BD +64106 & 5.43 $\pm$ 0.16 & 17.67 $\pm$ 0.62& 5.150 $\pm$ 0.098 & -83.26  $\pm$ 0.54  &  79.07\\
            &HD19820 & 4.46 $\pm$ 0.08 & 35.25 $\pm$ 0.46 & 4.562 $\pm$ 0.025 & -65.54 $\pm$ 0.16 & 79.21 \\
            &HD25443 & 4.74 $\pm$ 0.03 & 56.45 $\pm$ 0.33 & 4.734 $\pm$ 0.045 & -46.35 $\pm$ 0.28 & 77.20 \\
            &HD7927 & 3.22 $\pm$ 0.04 & 11.34 $\pm$ 0.66 & 3.026 $\pm$ 0.037 & -89.16 $\pm$ 0.35 & 79.50 \\
            &HD236633 & 5.58 $\pm$ 0.12 & 12.77 $\pm$ 0.62 & 5.576 $\pm$ 0.028 & -85.96 $\pm$ 0.15 & 80.27 \\
            \hline
            \textbf{Oct 22, 2022} & & & & & &\\
            \hline
            &HD19820 & 4.58 $\pm$ 0.15 & 36.60 $\pm$ 1.00 &  4.562 $\pm$ 0.025 & -65.54 $\pm$ 0.16 & 80.86 \\
            &BD+59 389 &  6.52 $\pm$ 0.15 & 19.35 $\pm$ 0.19 & 6.430 $\pm$ 0.022 & -81.86 $\pm$ 0.10 & 78.79\\
            \hline
            \textbf{Oct 23, 2022} & & & & & &\\
            \hline
            &BD +64106 & 5.55 $\pm$ 0.38 & 19.17 $\pm$ 0.37 & 5.150 $\pm$ 0.098 & -83.26  $\pm$ 0.54 & 77.57 \\
            &HD19820 & 4.56 $\pm$ 0.11 & 36.28 $\pm$ 0.99 & 4.562 $\pm$ 0.025 & -65.54 $\pm$ 0.16 & 78.18 \\
            &HD25443  & 4.84 $\pm$ 0.2 & 57.20 $\pm$ 1.04 &  4.734 $\pm$ 0.045 & -46.35 $\pm$ 0.28 & 76.45 \\
            &BD+59 389 & 6.58 $\pm$ 0.09 & 18.45 $\pm$ 0.50 & 6.430 $\pm$ 0.022 & -81.86 $\pm$ 0.10 & 79.69 \\
		\hline
	\end{tabular}
\end{table*}

\begin{table*}
	\centering
	\caption{Polarization measurements of unpolarized standard, HD14069 star at each epoch.}
	\label{tab:unpol}
        %\small
	\begin{tabular}{lccr} 
 \hline
		\hline
		Date & P $\pm$ $\sigma_P$ $(\%)$ & $\theta_{obs}$ $\pm$ $\sigma_{\theta}$ $(^\circ)$ & \\
    \hline
		\hline
             Oct 19, 2022 & 0.16 $\pm$ 0.01 & 20.14 $\pm$ 02.25 & \\
            Oct 21, 2022 & 0.13 $\pm$ 0.09 & 41.32 $\pm$ 19.93\\
            Oct 22, 2022 & 0.19 $\pm$ 0.08 & 43.13 $\pm$ 12.11 \\ 
            Oct 23, 2022 & 0.15 $\pm$ 0.10 & 85.45 $\pm$ 19.02 \\
		\hline
	\end{tabular}
\end{table*}

We restricted our analysis to stars with a G-band magnitude $<17$~mag, thus excluding fainter stars that typically exhibit low S/N and may yield unreliable results. Additionally, we only considered stars with a degree of polarization exceeding the 3$\sigma_p$ threshold.
 While these criteria led to the exclusion of many nearby foreground stars with low S/N in their polarization, we prioritized the selection of the most reliable measurements in our study, except in Sect.~\ref{sec:5.2.2}. After applying these two criteria, the remaining data comprises of 158, 151, 80, 161, and 88 stars towards Kronberger~1, Berkeley~69, Berkeley~71, King~8, and Berkeley~19, respectively.

\subsection{Archival data}\label{sec:3}
The observed polarization data is complemented with the imaging data from the Wide-Field Infrared Survey Explorer ( WISE; $\lambda$ = 3.4, 4.6, 12, 22 $\mu$m; resolutions = $6^{\prime\prime}.1$, $6^{\prime\prime}.4$, $6^{\prime\prime}.5$, $12^{\prime\prime}$; \citealt{WISE}) depicting the 2-dimensional distribution of warm dust of the ISM, and from the Herschel Infrared Galactic Plane Survey (Hi-GAL; $\lambda$ =  250, 350, 500 $\mu$m; resolutions =  $18^{\prime\prime}$, $25^{\prime\prime}$, $37^{\prime\prime}$; \citealt{HiGal}) for the cold dust distribution.

The polarization measurement of five additional clusters located in the same line of sight, NGC~2281 \citep{Eswaraiah2011}, NGC~1960 \citep{Eswaraiah2011}, Stock~8 \citep{Eswaraiah2011}, NGC~1931 \citep{Pandey2013}, and NGC~1893 \citep{Eswaraiah2011, Bijas2022} 
were used to study the large-scale dust distribution towards the anticenter direction. More details on these clusters are given in Table~\ref{tab:3}. 

\section{Results}\label{sec:4}
In this section, we highlight the polarization results obtained from the observations of five clusters toward the anticenter direction. 
The polarization measurements of 159, 152, 80, 161, and 88 stars towards Kronberger~1, Berkeley~69, Berkeley~71, King~8, and Berkeley~19 are added in the electronic media as extended tables.
The extended table lists the degree of polarization ($p_{obs}$), debiased polarization ($P$), polarization angle ($\theta$), Stokes parameters ($q$ and $u$), and their errors ($\sigma_p$, $\sigma_\theta$, $\sigma_q$, and $\sigma_u$) for the selected stars observed towards each individual cluster, labeled by the cluster name. The corresponding \textit{Gaia} DR3 source IDs of the observed stars are also appended to the tables. A sample of the observed polarization measurements is presented in Table~\ref{tab:obs}.
\begin{table*}
    \caption{A sample of polarization measurements of a few stars towards Kronberger~1 cluster. }.\label{tab:obs}
    \centering
    \begin{tabular}{r r r r r r r r r r r r}
\hline
\hline \\[-1.5ex]
  \multicolumn{1}{c}{Gaia ID} &
  \multicolumn{1}{c}{RAJ2000} &
  \multicolumn{1}{c}{DECJ2000} &
  \multicolumn{1}{c}{$p_{obs}$} &
  \multicolumn{1}{c}{$\sigma_p$} &
  \multicolumn{1}{c}{$P$} &
  \multicolumn{1}{c}{$\theta$} &
  \multicolumn{1}{c}{$\sigma_{\theta}$} &
  \multicolumn{1}{c}{$q$} &
  \multicolumn{1}{c}{$\sigma_q$} &
  \multicolumn{1}{c}{$u$} &
  \multicolumn{1}{c}{$\sigma_u$} \\ \\[-1.5ex]
  \multicolumn{1}{c}{} &
  \multicolumn{1}{c}{(deg)} &
  \multicolumn{1}{c}{(deg)} &
  \multicolumn{1}{c}{$(\%)$} &
  \multicolumn{1}{c}{$(\%)$} &
  \multicolumn{1}{c}{$(\%)$} &
  \multicolumn{1}{c}{(deg)} &
  \multicolumn{1}{c}{(deg)} &
  \multicolumn{1}{c}{$(\%)$} &
  \multicolumn{1}{c}{$(\%)$} &
  \multicolumn{1}{c}{$(\%)$} &
  \multicolumn{1}{c}{$(\%)$}\\ \\[-1.5ex]
\hline \\[-1.5ex]
182856896896621952 & 81.96833 & 34.78284 & 2.88 & 0.15 & 2.88 & -12.8 & 1.6 & 2.592 & 0.142 & -1.246 & 0.164\\
  182856931256349184 & 81.97639 & 34.79769 & 2.30 & 0.37 & 2.27 & -15.1 & 4.7 & 1.985 & 0.367 & -1.154 & 0.375\\
  182857309213463936 & 81.97693 & 34.81212 & 2.48 & 0.08 & 2.48 & -06.2 & 1.3 & 2.423 & 0.081 & -0.529 & 0.113\\
  182856725097922816 & 81.97730 & 34.79082 & 3.02 & 0.10 & 3.02 & -25.0 & 0.9 & 1.946 & 0.081 & -2.314 & 0.113\\
  182856725097925120 & 81.97839 & 34.78405 & 3.74 & 0.32 & 3.73 & -33.2 & 2.4 & 1.491 & 0.312 & -3.427 & 0.322\\
  182857515371880576 & 81.98116 & 34.83915 & 3.96 & 0.18 & 3.96 & -11.0 & 1.4 & 3.668 & 0.180 & -1.488 & 0.198\\
  182855866104483712 & 81.98147 & 34.75075 & 3.11 & 0.11 & 3.11 & -19.7 & 1.1 & 2.398 & 0.096 & -1.973 & 0.125\\
  182856725097921920 & 81.98307 & 34.78822 & 2.45 & 0.15 & 2.45 & -19.0 & 1.8 & 1.931 & 0.137 & -1.514 & 0.159\\
  182857412292669824 & 81.98512 & 34.82646 & 2.46 & 0.10 & 2.46 & -06.6 & 1.4 & 2.398 & 0.097 & -0.560 & 0.125\\
  182857511072807168 & 81.98736 & 34.84172 & 4.65 & 0.70 & 4.6 & -18.1 & 4.3 & 3.751 & 0.700 & -2.754 & 0.708\\
  182857511072807296 & 81.98770 & 34.84010 & 4.56 & 1.29 & 4.37 & -11.4 & 8.1 & 4.202 & 1.289 & -1.760 & 1.296\\
  182855866104477184 & 81.98849 & 34.75599 & 3.28 & 0.08 & 3.28 & -05.5 & 1.0 & 3.217 & 0.081 & -0.625 & 0.114\\
  182857309213462528 & 81.99054 & 34.80572 & 4.95 & 0.67 & 4.9 & -15.2 & 3.9 & 4.276 & 0.665 & -2.499 & 0.671\\
  182855866104483968 & 81.99228 & 34.74149 & 3.57 & 0.22 & 3.56 & -08.4 & 1.9 & 3.420 & 0.222 & -1.032 & 0.236\\
  182857446652403968 & 81.99337 & 34.82917 & 5.28 & 0.10 & 5.28 & -24.6 & 0.5 & 3.457 & 0.081 & -3.991 & 0.114\\
  182855866104481280 & 81.99345 & 34.74653 & 4.09 & 0.09 & 4.09 & -11.8 & 0.8 & 3.745 & 0.081 & -1.632 & 0.113\\
  182857103055039232 & 81.99372 & 34.78785 & 2.48 & 0.15 & 2.48 & -11.8 & 1.9 & 2.274 & 0.147 & -0.989 & 0.167\\
  182855866104476928 & 81.99543 & 34.75104 & 3.09 & 0.09 & 3.09 & -17.8 & 1.0 & 2.507 & 0.081 & -1.800 & 0.113\\
  182855900465355392 & 81.99757 & 34.75706 & 3.16 & 0.17 & 3.16 & -15.9 & 1.6 & 2.688 & 0.160 & -1.661 & 0.179\\
  182857343573191808 & 81.99786 & 34.81937 & 3.47 & 0.09 & 3.47 & -15.2 & 0.9 & 2.993 & 0.083 & -1.749 & 0.115\\
  182857549731606656 & 82.00080 & 34.84838 & 1.73 & 0.13 & 1.73 & -06.3 & 2.5 & 1.689 & 0.128 & -0.379 & 0.151\\
  182856274125658624 & 82.00632 & 34.77641 & 2.78 & 0.34 & 2.76 & -16.1 & 3.5 & 2.356 & 0.339 & -1.483 & 0.347\\
  182857442353248256 & 82.01029 & 34.82783 & 1.34 & 0.11 & 1.34 & -01.2 & 2.9 & 1.340 & 0.110 & -0.057 & 0.136\\
  182857137414762624 & 82.01218 & 34.80639 & 2.98 & 0.09 & 2.98 & -12.8 & 1.1 & 2.682 & 0.083 & -1.289 & 0.115\\
  182857481012123136 & 82.01261 & 34.85074 & 5.23 & 0.11 & 5.23 & -27.9 & 0.5 & 2.934 & 0.082 & -4.328 & 0.115\\
  182863425247385344 & 82.01404 & 34.86500 & 5.62 & 1.04 & 5.52 & 02.7 & 5.3 & 5.594 & 1.042 & 0.519 & 1.049\\
  182855831744729088 & 82.01743 & 34.75095 & 2.99 & 0.10 & 2.99 & -20.7 & 1.0 & 2.239 & 0.087 & -1.980 & 0.118\\
  182856274125655296 & 82.01784 & 34.77587 & 1.66 & 0.09 & 1.66 & -07.6 & 2.1 & 1.605 & 0.091 & -0.435 & 0.121\\
  182857270554609024 & 82.01993 & 34.82942 & 4.06 & 0.09 & 4.06 & -16.7 & 0.7 & 3.392 & 0.081 & -2.233 & 0.114\\
 \\[-1.5ex]
  \hline
  \end{tabular}
  \tablefoot{The measured degree of polarization ($P_{obs}$), its debiased value ($P$), polarization angle ($\theta$), Stokes parameters $q$ and $u$, and their respective uncertainties are tabulated along with their corresponding \textit{Gaia} DR3 ID and sky coordinates (RAJ2000, and DECJ2000)}
  \end{table*}

\subsection{Polarization Sky map}\label{sec:4.1}
\begin{figure*}
\centering
   \includegraphics[width=17cm]{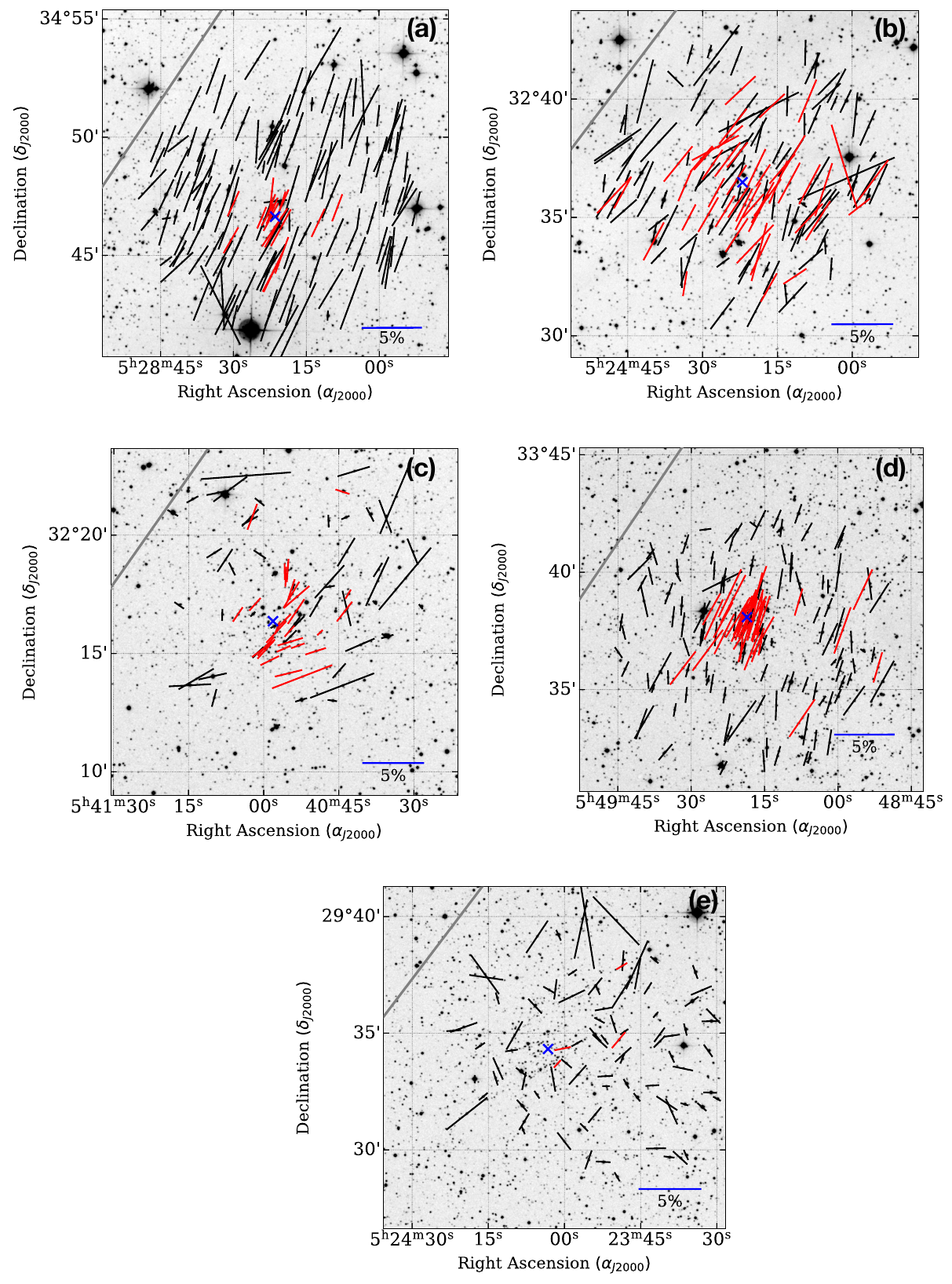}
     \caption{Polarization line segments overlaid on DSS R-band $25^{\prime} \times 25^\prime$ field towards Kronberger~1 in panel (a), Berkeley~69 in (b), Berkeley~71 in (c), King~8 in (d), and Berkeley~19 in panel (e). {The length of the segment is proportional to the degree of polarization, and orientation depicts the polarization angle.} The red lines represent the polarization measurements of probable member stars, and the solid gray line {in the upper-left corner} corresponds to the orientation of the Galactic plane. {A reference line of $5\%$ polarization and $90^\circ$ polarization angle is also drawn in the lower-right corner of each panel. }}
     \label{fig:Fig1}
\end{figure*}

The sky projection of the polarization segments of all the observed stars towards different cluster directions is overlaid on $25^\prime \times 25^\prime$ DSS\footnote{\href{https://irsa.ipac.caltech.edu/data/DSS/}{https://irsa.ipac.caltech.edu/data/DSS/}} R-band images (in Equatorial coordinates) and shown in Fig.~\ref{fig:Fig1}(a) to Fig.~\ref{fig:Fig1}(e). The red line segments indicate the member stars with a membership probability of more than 50\%, as calculated in \citet{Hunt2023}. The length of each segment is proportional to the degree of polarization while their orientation traces the measured optical polarization position angle ($\theta$, Eq.~\ref{eq:6}).
A reference line segment of 5\% polarization and $90^\circ$ polarization angle is drawn in the lower right corner of each panel. The solid gray line in the upper-left corner of each figure depicts the orientation of the Galactic plane (constant latitude). 

%\begin{figure*}
%\centering
%\subfigure[Kronberger 1]
%{\label{fig:1a}\includegraphics[width=0.40\textwidth]{revised2/Fig1_R2_Kron1.pdf}}
%\subfigure[Berkeley 69]{\label{fig:1b}\includegraphics[width=0.40\textwidth]{revised2/Fig1_R2_Be69.pdf}}
%\subfigure[Berkeley 71]{\label{fig:1c}\includegraphics[width=0.40\textwidth]{revised2/Fig1_R2_Be71.pdf}}
%\subfigure[King8]{\label{fig:1d}\includegraphics[width=0.40\textwidth]{revised2/Fig1_R2_King8.pdf}}
%\subfigure[Berkeley 19]{\label{fig:1e}\includegraphics[width=0.40\textwidth]{revised2/Fig1_R2_Be19.pdf}}

%\caption{Polarization line segments overlaid on DSS R-band $25^{\prime} \times 25^\prime$ field towards Kronberger~1 in panel (a), Berkeley~69 in (b), Berkeley~71 in (c), King~8 in (d), and Berkeley~19 in panel (e). {The length of the segment is proportional to the degree of polarization, and orientation depicts the polarization angle.} The red lines represent the polarization measurements of probable member stars, and the solid gray line {in the upper-left corner} corresponds to the orientation of the Galactic plane. {A reference line of $5\%$ polarization and $90^\circ$ polarization angle is also drawn in the lower-right corner of each panel. }}\label{fig:Fig1}
%\end{figure*}

\subsection{Distribution of stars in $P-\theta$}\label{sec:4.2}
\begin{figure*}
\centering
   \includegraphics[width=17cm]{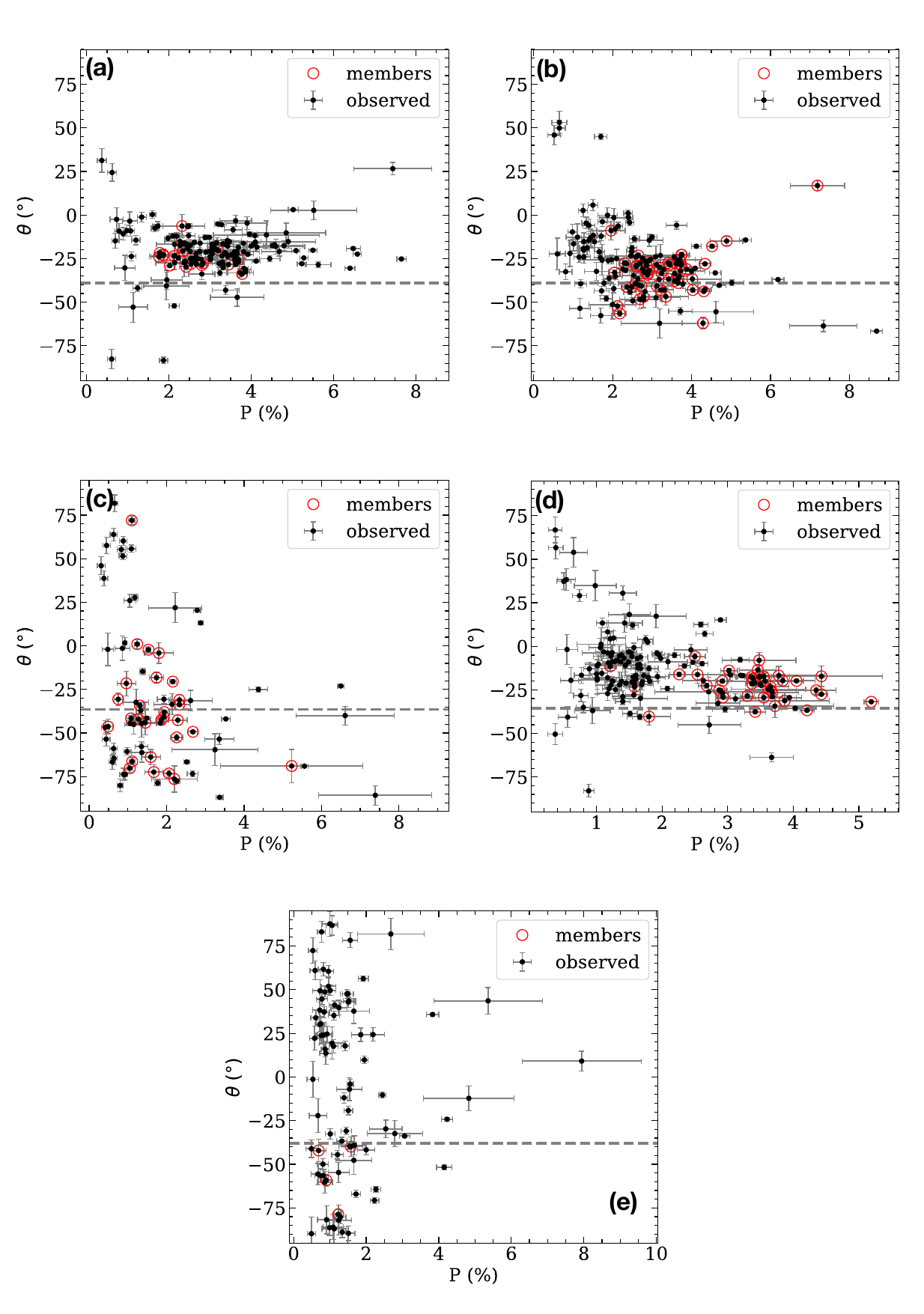}
   \caption{Scatter plot of $P$ and $\theta$ of observed stars towards Kronberger~1 (panel a), Berkeley~69 (panel b), Berkeley~71 (panel c), King~8 (panel d), and Berkeley~19 (panel e)  with errorbars representing uncertainties in the polarization measurements ($\sigma_P$ and $\sigma_\theta$). The red open circle in all the panels corresponds to the $P$ and $\theta$ values of probable member stars. The gray dashed line depicts the orientation of the Galactic plane towards the respective cluster direction.}
    \label{fig:Fig2}
\end{figure*}

A detailed examination of polarization measurements towards the observed clusters is presented in Fig.~\ref{fig:Fig2}(a) to Fig.~\ref{fig:Fig2}(e).
%\begin{figure*}
%\centering
%\subfigure[Kronberger 1]{\label{fig:2a}\includegraphics[width=0.42\textwidth]{revised2/Fig2_R2_Kron1.pdf}}
%\subfigure[Berkeley 69]{\label{fig:2b}\includegraphics[width=0.42\textwidth]{revised2/Fig2_R2_Be69.pdf}}
%\subfigure[Berkeley 71]{\label{fig:2c}\includegraphics[width=0.42\textwidth]{revised2/Fig2_R2_Be71.pdf}}
%\subfigure[King8]{\label{fig:2d}\includegraphics[width=0.42\textwidth]{revised2/Fig2_R2_King8.pdf}}
%\subfigure[Berkeley 19]{\label{fig:2e}\includegraphics[width=0.42\textwidth]{revised2/Fig2_R2_Be19.pdf}}
%\caption{Scatter plot of $P$ and $\theta$ of observed stars towards Kronberger~1 (panel a), Berkeley~69 (panel b), Berkeley~71 (panel c), King~8 (panel d), and Berkeley~19 (panel e)  with errorbars representing uncertainties in the polarization measurements ($\sigma_P$ and $\sigma_\theta$). The red open circle in all the panels corresponds to the $P$ and $\theta$ values of probable member stars. The gray dashed line depicts the orientation of the Galactic plane towards the respective cluster direction.}
%    \label{fig:Fig2}
%\end{figure*}
The member stars having membership probability more than 50\% are denoted by the red open circles. The orientation of the Galactic plane towards each cluster direction is represented by a gray dashed line. In all the panels of Fig.~\ref{fig:Fig2}, we notice that the stars exhibiting a low degree of polarization show large dispersion in position angle. 
The spread in the polarization angle reduced with increasing polarization. 

In the case of Kronberger~1 (Fig.~\ref{fig:Fig2}a), most of the member stars exhibit uniform polarization angles ($\sim  -27^\circ$) with a small dispersion of
$\sim 5^\circ$. Only one member star, with a membership probability of $90\%$, exhibits a larger
polarization angle ($= - 6^\circ$) in contrast to the remaining detected member stars.
The weighted average degree of polarization for the Kronberger~1 cluster is calculated to be $2.55 \pm 0.57 ~\%$ and the polarization angle of $ \sim -27^\circ$, having a standard deviation of $5^\circ$ among the member stars. 
This calculation is performed using the following equations:
\begin{equation}\label{wtAv}
    \Bar{X} = \frac{\sum_{i} w_i~X_i}{\sum_{i} w_i}\; ,
\end{equation}
and 
\begin{equation}\label{wtsd}
    \sigma_{\Bar{X}} = \sqrt{ \frac{\sum_i w_i (X_i - \Bar{X})^2}{\frac{N-1}{N} \sum_i w_i } }~.
\end{equation}
Here, $X$ denotes $P$ values for $N$ number of stars, and we used the weighted circular mean from the circular statistics package of \textit{Astropy} in the range $(0, \pi)$ to calculate the mean polarization angle.
To perform the analyses and present the results, we choose to use the $P$ and $\theta$ values as they directly provide information on the amplitude of the polarization source field and on the plane-of-sky orientation of the magnetic field, which is associated with dust clouds. We verified that running our analysis in ($q,\,u$) space and converting to ($P,\,\theta$) space led to the same results. In our calculation the weights ($w_i$) were defined from corresponding uncertainties ($\sigma = \sigma_P$ or $\sigma_{\theta}$ ) as
\begin{equation}\label{wt}
   w_{i} = \frac{1}{\sigma_i^2} ~. 
\end{equation}
Furthermore, it is seen in Fig.~\ref{fig:Fig2}(a) that the non-member stars towards Kronberger~1 also display nearly similar polarization angles as that of the member stars.

In contrast, the distribution of stars towards Berkeley~69 (Fig.~\ref{fig:Fig2}b) in $P-\theta$ plane roughly denotes three groups.
The first group consists of four stars with a low degree of polarization and polarization angles around $50^\circ$.
The second group exhibits a slightly higher degree of polarization (approximately $1.5\%$) and a polarization angle of $\theta \approx -11^\circ$.
The third group primarily consists of all the member stars, showing a higher average degree of polarization (around $3 \%$) compared to the other groups. These stars also have a lower polarization angle (approximately $-33^\circ$) compared to the stars in the second group. The differences in polarization among these three groups suggest the presence of variations in the magnetized ISM, either along the line of sight or in the plane of the sky. The weighted average degree of polarization and weighted circular mean polarization angle with their respective weighted standard deviation for member stars are derived using Eq.~\ref{wtAv}, Eq.~\ref{wtsd} and Eq.~\ref{wt}, yielding values approximately equal to $3.23 \pm 0.68~\%$ and $ -34^\circ \pm 8^\circ$.
Similar to Berkeley~69, the stars towards the King~8 (Fig.~\ref{fig:Fig2}d) cluster also show distinct grouping in the $P-\theta$ plane. The weighted average degree of polarization of member stars towards King~8 is $3.13 \pm 0.92$, with a polarization angle of $-26^\circ \pm 7^\circ$. 

The stars towards Berkeley~71 (Fig.~\ref{fig:Fig2}c) and Berkeley~19 (Fig.~\ref{fig:Fig2}e) display a low degree of polarization and large dispersion in polarization angle. The weighted mean polarization for Berkeley~71 is $1.50 \pm 0.53~\%$, and the polarization angle is around $-40^\circ \pm 16^\circ$. Similarly, the weighted mean degree of polarization for Berkeley~19 member stars is calculated to be $1.08 \pm 0.27~ \%$ with a polarization angle of $-62^\circ \pm 15^\circ$. In most cases, the polarization angle tracing the plane-of-sky orientation of the magnetic field is closely aligned with that of the orientation of the Galactic plane, implying the dust is relaxed in those directions except for stars towards Berkeley~71 and Berkeley~19. The polarization angles of only a few member stars in Berkeley~71 exhibit alignment with the Galactic plane, while most of the field stars and cluster members display a scattered orientation. This suggests that the magnetic field in this direction exhibits inhomogeneity in small-scale regions.
Furthermore, member stars in this direction show a helical pattern in the polarization angle on the sky plane, as seen in Fig.~\ref{fig:Fig1}(c). This is likely the result of inhomogeneities in the foreground clouds as discussed in Sect.~\ref{sec:5.1} and Sect.~\ref{sec:5.2} where we show that the polarization segments closely align with the morphology of the foreground cloud. Berkeley~19, on the other hand, is the most distant cluster in our observation campaign. Most of the observed stars in this direction are foreground to the cluster, suffering less dust extinction. Consequently, they exhibit a low degree of polarization with a large scatter in polarization angle.

\section{Discussions}\label{sec:5}
In this section, we present discussions about the spatial and line-of-sight dust distribution towards individual clusters by combining the polarization measurements with the archival data, including WISE, \textit{Herschel}, and \textit{Gaia}.

\subsection{Spatial dust distribution}\label{sec:5.1}
We have seen in Sect.~\ref{sec:4.2} that stars towards Berkeley~71 and Berkeley~19 show large dispersion in polarization angle. Stars towards the other clusters either show a nearly uniform polarization or some grouping in the polarization measurements. The variation in the degree of polarization and polarization angle could be linked with the spatial distribution of foreground dust clouds along with the plane-of-sky magnetic field orientation. A uniform foreground dust distribution with a uniform plane-of-sky magnetic field is expected to show low dispersion in the polarization angle. However, large dispersion in the polarization angle indicates the presence of non-uniform dust distribution or varying magnetic field orientation on the sky plane.  
In order to infer such a possible scenario, we analyzed two-dimensional maps of warm and cold dust distribution in the following subsections.

\subsubsection{WISE maps}\label{sec:5.1.1}
\begin{figure*}
\centering
   \includegraphics[width=17cm]{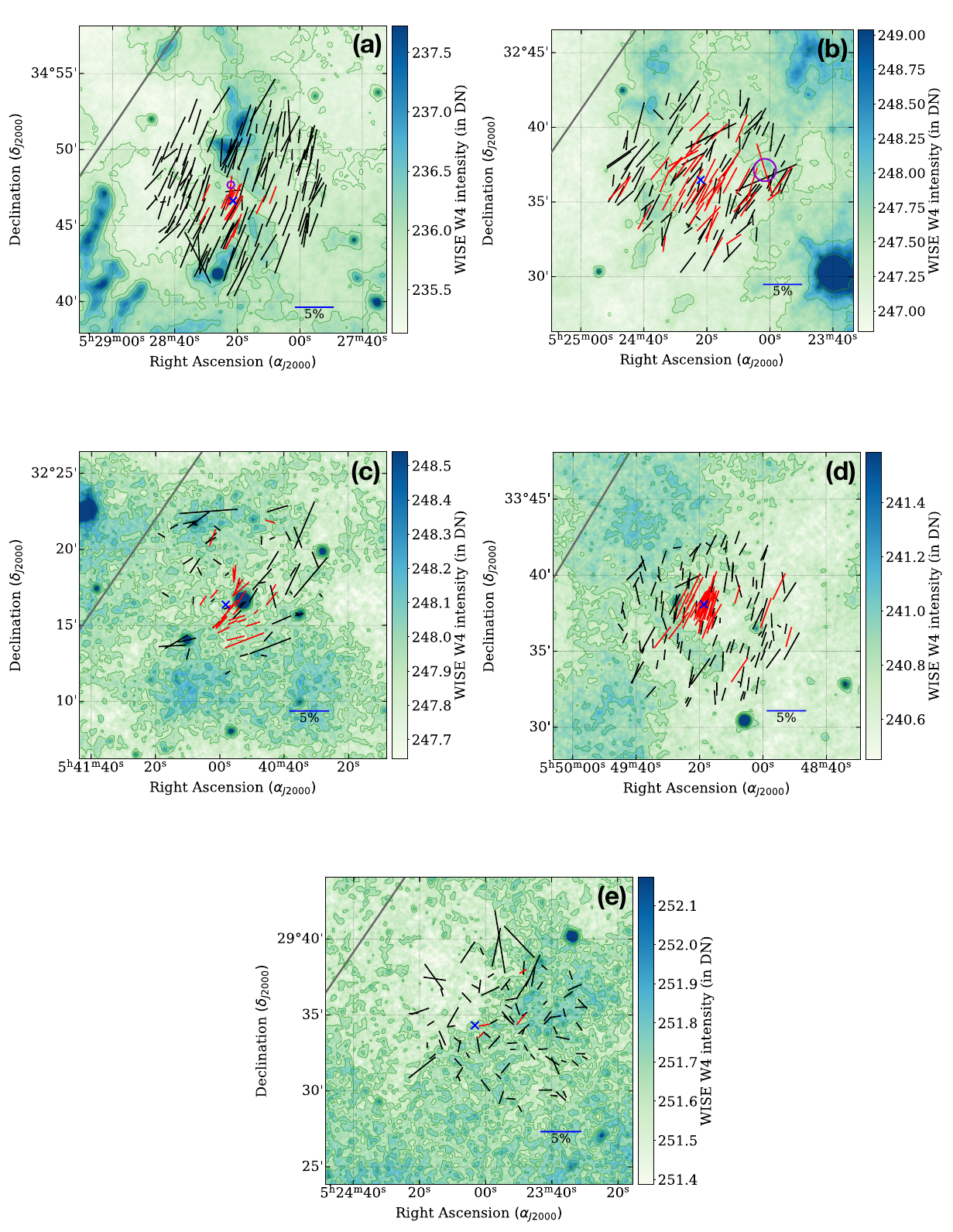}
    \caption{WISE W4 map of $25^{\prime} \times 25^{\prime}$ fields, centered at Kronberger~1 (panel a), Berkeley~69 (panel b), Berkeley~71 (panel c), King~8 (panel d), and Berkeley~19 (panel e). The polarization segments are the same as in Fig.~\ref{fig:Fig1} for member stars (in red), and field stars (in black) are superimposed on the respective WISE map. The color scale in the maps represents the WISE W4 intensity (in DN), and contours correspond to its constant value.}  \label{fig:Fig3}
\end{figure*}

The intensity images in WISE W4 ($22 \, \mu$m) \citep{WISEW4} depict the line-of-sight integrated distribution of warm dust across the plane of the sky. Figure~\ref{fig:Fig3}(a) to Fig.~\ref{fig:Fig3}(e) show the polarization measurements towards each cluster direction superimposed on WISE W4 maps covering a $25^\prime$ field centered at the cluster location. 
%\begin{figure*}
%\centering
%\subfigure[Kronberger 1]{\label{fig:3a}\includegraphics[width=0.42\textwidth]{revised2/Fig3_R2_Kron1.pdf}}
%\subfigure[Berkeley 69]{\label{fig:3b}\includegraphics[width=0.42\textwidth]{revised2/Fig3_R3_Be69.pdf}}
%\subfigure[Berkeley 71]{\label{fig:3c}\includegraphics[width=0.42\textwidth]{revised2/Fig3_R2_Be71.pdf}}
%\subfigure[King8]{\label{fig:3d}\includegraphics[width=0.42\textwidth]{revised2/Fig3_R2_King8.pdf}}
%\subfigure[Berkeley 19]{\label{fig:3e}\includegraphics[width=0.42\textwidth]{revised2/Fig3_R2_Be19.pdf}}
%\caption{WISE W4 map of $25^{\prime} \times 25^{\prime}$ fields, centered at Kronberger~1 (panel a), Berkeley~69 (panel b), Berkeley~71 (panel c), King~8 (panel d), and Berkeley~19 (panel e). The polarization segments are the same as in Fig.~\ref{fig:Fig1} for member stars (in red), and field stars (in black) are superimposed on the respective WISE map. The color scale in the maps represents the WISE W4 intensity (in DN), and contours correspond to its constant value.}  \label{fig:Fig3}
%\end{figure*}
The color coding and orientation of the polarization vectors align with those in Fig.~\ref{fig:Fig1}. The color bar located at the right side of the panels of Fig.~\ref{fig:Fig3} corresponds to the WISE W4 intensity in digital numbers (DN), and the green contours denote constant W4 intensity contours. The color gradient seen in these panels suggests a patchy spatial distribution of the dust towards corresponding cluster directions.
In the direction towards the Kronberger~1 cluster, 
the polarization angle of the majority of the stars is oriented systematically away from the Galactic plane (see Fig.~\ref{fig:Fig2}a and Fig.~\ref{fig:Fig3}a). A closer inspection of Fig.~\ref{fig:Fig3}(a) reveals that the orientation of the polarization segments of a few stars seems to follow the curvature of the filament seen in the WISE map. 
However, we require more polarization data of stars located at similar distances, covering a wider area beyond the whole filament, to confirm the relation of the polarization features with the large-scale filament geometry. 
Towards this direction, the polarization angle of most member stars appears to align with each other, except for one outlier (shown in the purple circle in Fig.~\ref{fig:Fig3}a), having a polarization angle of $-6^\circ$.

In the case of Berkeley~69, most of the member stars present in group 3 (see Sect.~\ref{sec:4.2} and Fig.~\ref{fig:Fig2}b) show parallel alignment with the Galactic plane except for one star. This star exhibits a degree of polarization of $7.22 \pm 0.69 \%$ and a polarization angle of $15^\circ \pm 3^\circ$. The considerable deviation in $P$ and $\theta$, or $q$ and $u$ values of this star from the remaining member stars, raises doubts about its membership status or suggests the presence of intrinsic polarization. The sky position of this star is at the edge of the cluster, indicating the possibility of inhomogeneous dust distribution at the outskirts of the cluster. We plotted the polarization vectors on the WISE W4 image in Fig.~\ref{fig:Fig3}(b) to investigate this further. The figure clearly illustrates a non-uniform distribution of dust in the sky region towards the cluster direction. Interestingly, the central region appears to have less dust compared to the periphery of the cluster, as traced by WISE W4.  Here, the star with the polarization value of $7.22\%$ is denoted by a purple circle. The higher polarization could be attributed to the presence of warm dust, as suggested by the increased WISE W4 intensity at the star's location in Fig.~\ref{fig:Fig3}(b). The large difference in angle could, however, indicate that the star is intrinsically polarized.  We require multi-band polarization data to confirm the possibility of intrinsic polarization. Excluding this star, the weighted average of the degree of polarization and polarization angle of the cluster is estimated to be $3.23\% \pm 0.68\%$ and $-33.9^\circ \pm 8.2^\circ$.  

The dust traced by WISE W4 appears to be nearly uniformly distributed toward Berkeley~71 (Fig.~\ref{fig:Fig3}c) and Berkeley~19 (Fig.~\ref{fig:Fig3}e), as seen by WISE W4 contours. Consequently, the significant spread in the polarization angle cannot be solely attributed to the presence of warm dust. In contrast, Fig.~\ref{fig:Fig3}(d) reveals that the dust traced by WISE W4 towards King~8 is less concentrated toward the center of the cluster than in the peripheral regions. As a result, the member stars located closer to the cluster center exhibit distinct polarization behavior compared to field stars dispersed towards the edge of the cluster.

\subsubsection{Herschel maps}\label{sec:5.1.2}
In order to examine the distribution of cold dust and their correspondence in polarization measurements, we searched for our observed locations in Herschel High-Level Images \citep[HHLI;][]{HHLI} and found the data corresponding to only two cluster directions, i.e., Kronberger~1, and Berkeley~71.
\begin{figure}
  \resizebox{\hsize}{!}{\includegraphics{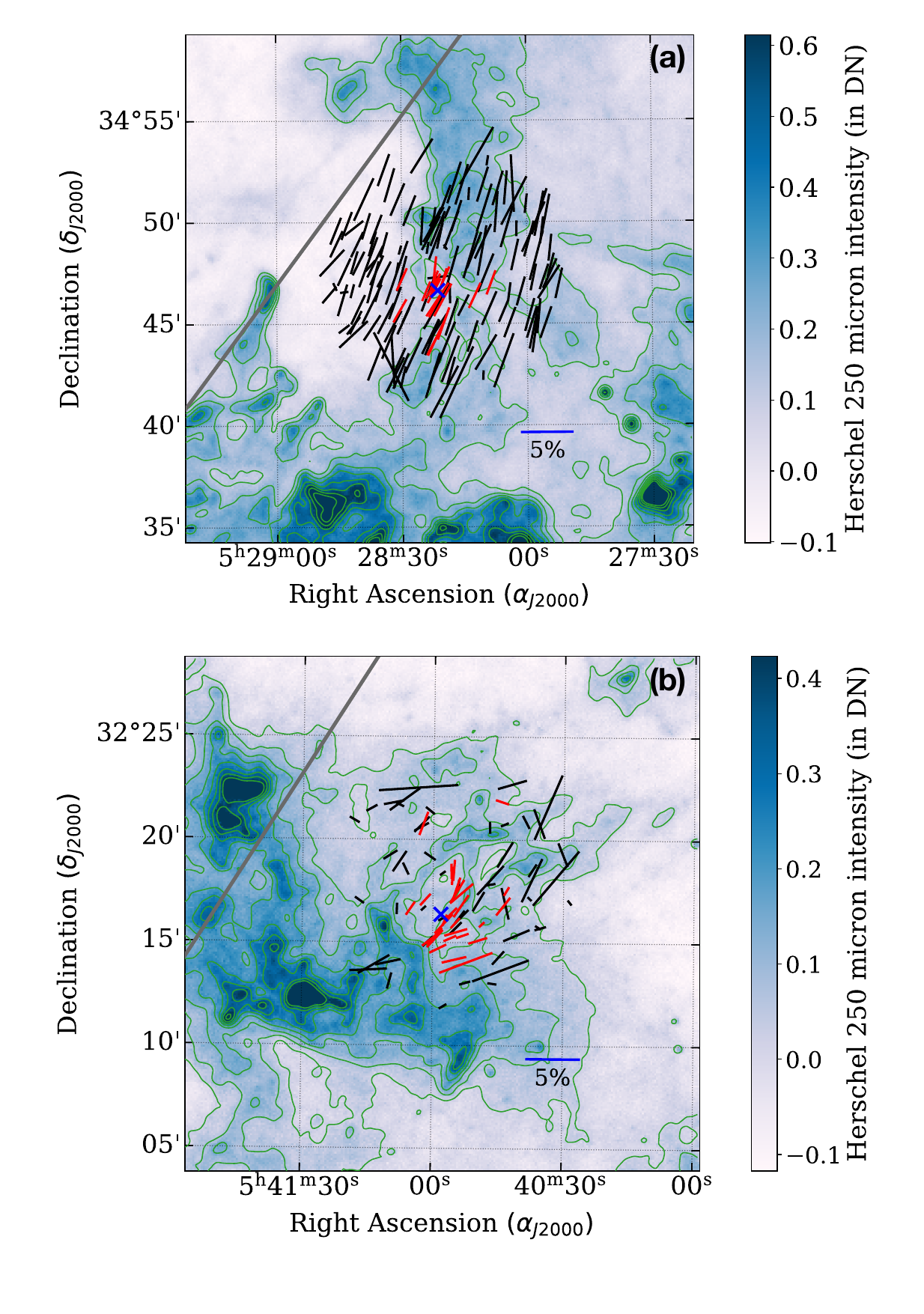}}
  \caption{Herschel 250 $\mu m$ dust map of Kronberger 1, in Panel (a) and Berkeley 71 in Panel (b). The contours correspond to the constant Herschel 250 $\mu m$ intensity. The polarization measurements are over-plotted on the dust map with the color-coding and orientation similar to Fig.~\ref{fig:Fig1}. A $5\%$ polarization line perpendicular to the celestial North is also added at the bottom-right side.}
  \label{fig:Fig4}
\end{figure}
%\begin{figure*}
%\centering
%\subfigure[Kronberger 1]{\label{fig:4a}\includegraphics[width=0.42\textwidth]{revised2/Fig4a_R2_Kron1.pdf}}
%\subfigure[Berkeley 71]{\label{fig:4b}\includegraphics[width=0.42\textwidth]{revised2/Fig4b_R2_Be71.pdf}}
%\caption{Herschel 250 $\mu m$ dust map of Kronberger 1, in Panel (a) and Berkeley 71 in Panel (b). The contours correspond to the constant Herschel 250 $\mu m$ intensity. The polarization measurements are over-plotted on the dust map with the color-coding and orientation similar to Fig.~\ref{fig:Fig1}. A $5\%$ polarization line perpendicular to the celestial North is also added at the bottom-right side.}\label{fig:Fig4}
%\end{figure*}
The color gradient and the contours in Fig.~\ref{fig:Fig4} depict the Herschel 250 $\mu$m intensity map plotted in equatorial coordinates. Similar to the WISE W4 map (Fig.~\ref{fig:Fig3}a), the Herschel 250 $\mu m$ map towards Kronberger~1 
(Fig.~\ref{fig:Fig4}a) also exhibits a filamentary feature where the polarization angle seems to be parallel to its longer axis. However, in the case of Berkeley~71, unlike the WISE W4 image (Fig.~\ref{fig:Fig3}c), the Hershel 250 $\mu$m map displayed in Fig.~\ref{fig:Fig4}(b), has some filamentary loop features.
Notably, the polarization vectors of the member stars appear to align with the 250-$\mu$m dust loop features, particularly in the vicinity of the cluster center. This correlation between the inhomogeneous cold dust emission and the polarization explains the significant scatter observed in the polarization angles towards the Berkeley~71 cluster.
\subsection{Line of sight dust distribution}\label{sec:5.2}
The estimation of the number of dust clouds and their distance along the line of sight is important for understanding the 3D distribution of dust and, hence, the structure of the Galaxy. To determine the distance of foreground dust layers towards each cluster direction, we attempted to employ kinematic distance measurements based on spectral information obtained from the HI4PI survey \citep{HI4PI} for the diffuse neutral medium, as well as molecular emission data from the $^{12}CO$ survey by \citep{Dame2001}. However, the kinematic distances obtained for the anticenter direction using the A5 model following the method outlined by \citet{Reid2014parameters} yielded unreliable values with large uncertainties. This discrepancy is attributed to the fact that, towards the anticenter direction, the major velocity component of the dust cloud is oriented along the plane of the sky rather than in the line of sight, following a circular rotation model. Consequently, the small radial velocity component and associated uncertainties contribute to significant uncertainties in the derived kinematic distances.

Nevertheless, as discussed in Sect.~\ref{sec:1}, the polarization of background starlight and extinction combined with distance information provides an alternative approach to infer the dust distribution along the line of sight.
The degree of polarization and extinction is anticipated to exhibit a jump when starlight encounters a dust layer. The polarization jump is also accompanied by a change in the polarization angle if the dust clouds along the line of sight are permeated by magnetic field lines with different orientations.
Consequently, the variation of polarization and extinction as a function of distance serves as a proxy for the dust distribution along the line of sight. This section provides detailed analyses for determining the number of dust layers present along the line of sight towards each cluster direction. It utilizes polarization information with distance through qualitative and quantitative methods and critically examines the constraints inherent in the applied methodologies.
 
\subsubsection{Visual inspection}\label{sec:5.2.1}
\begin{figure*}
\centering
\subfigure[]
{\label{fig:5a}\includegraphics[width=0.44\textwidth]{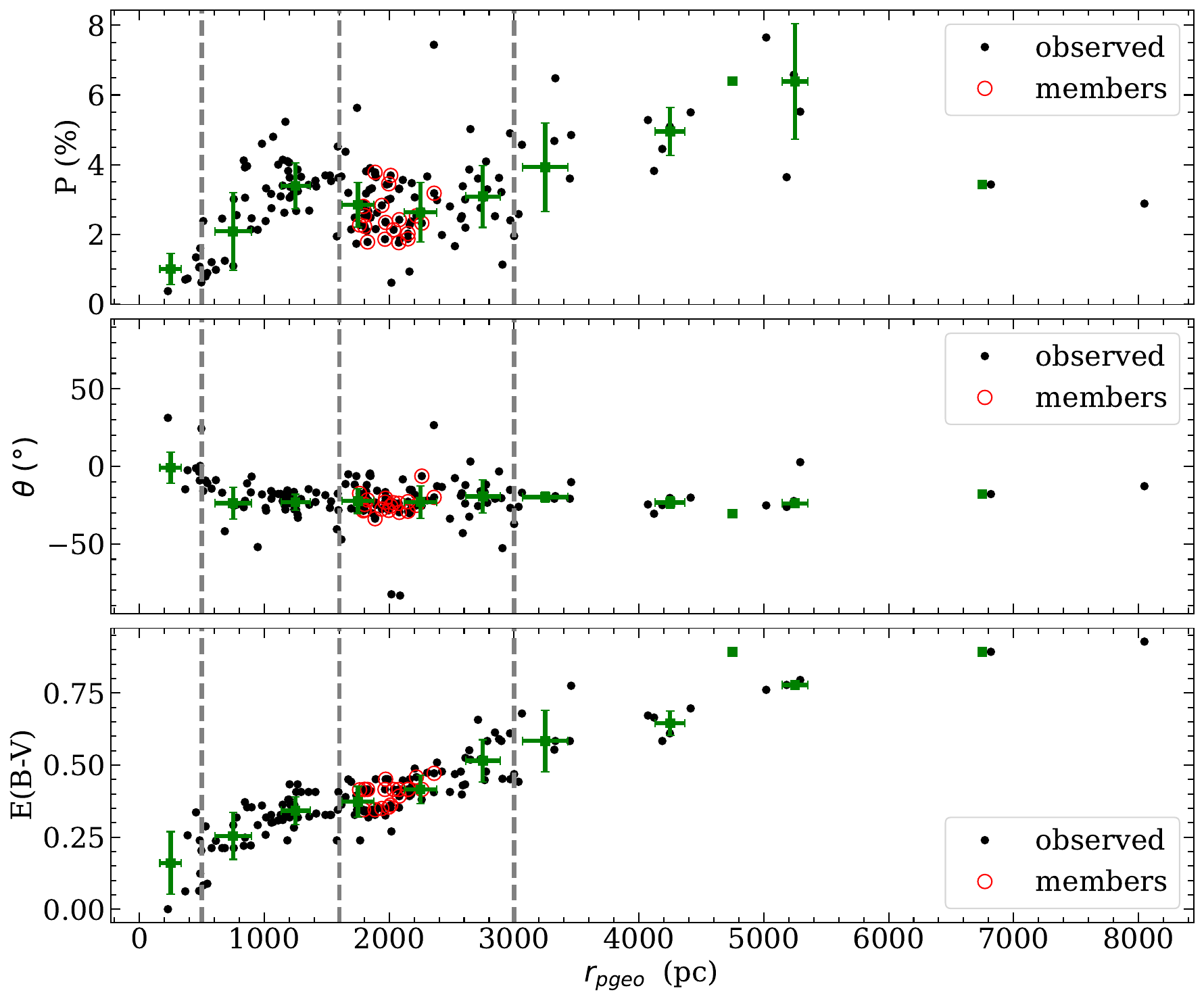}}
\subfigure[]
{\label{fig:5b}\includegraphics[width=0.44\textwidth]{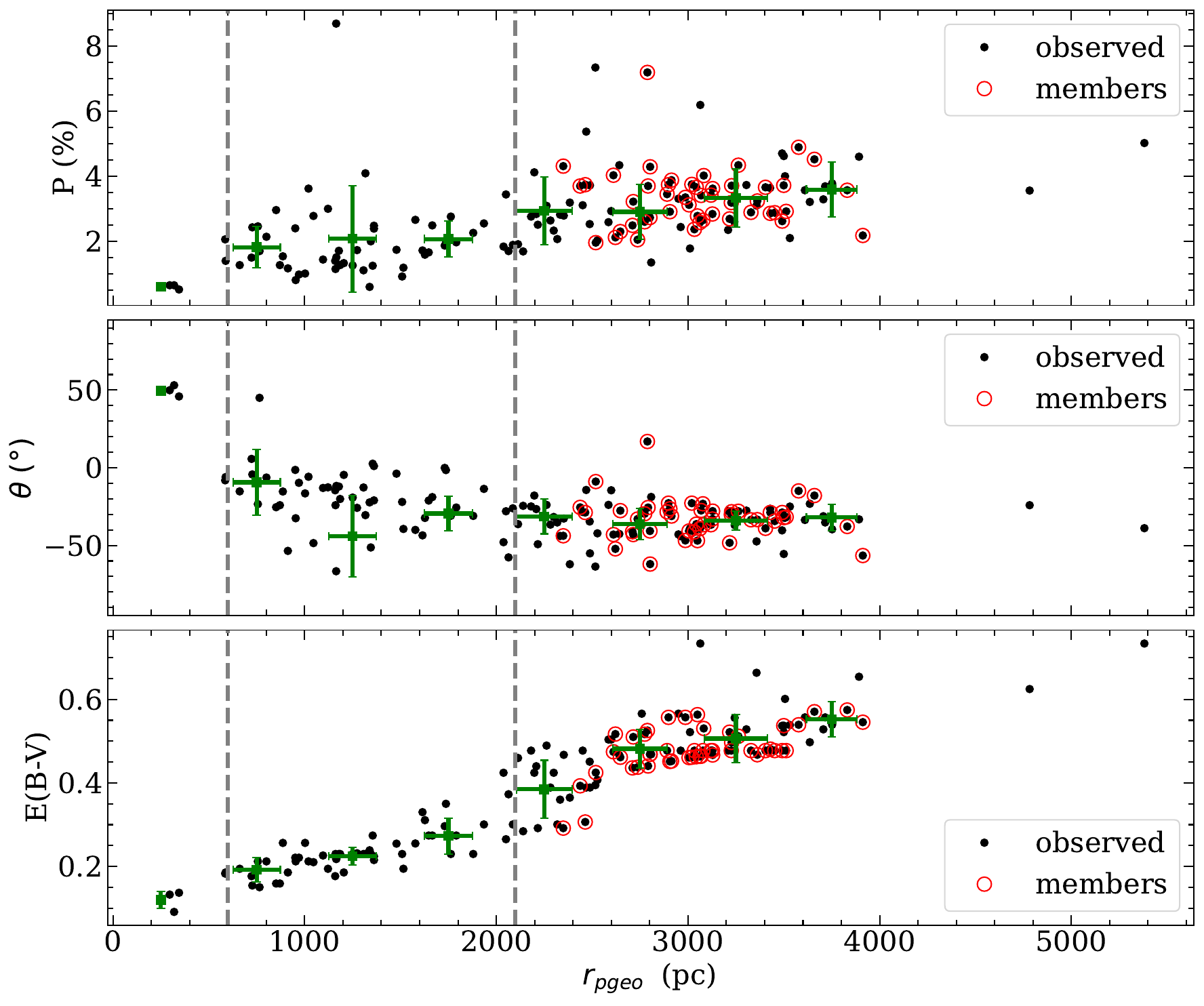}}
\subfigure[]
{\label{fig:5c}\includegraphics[width=0.44\textwidth]{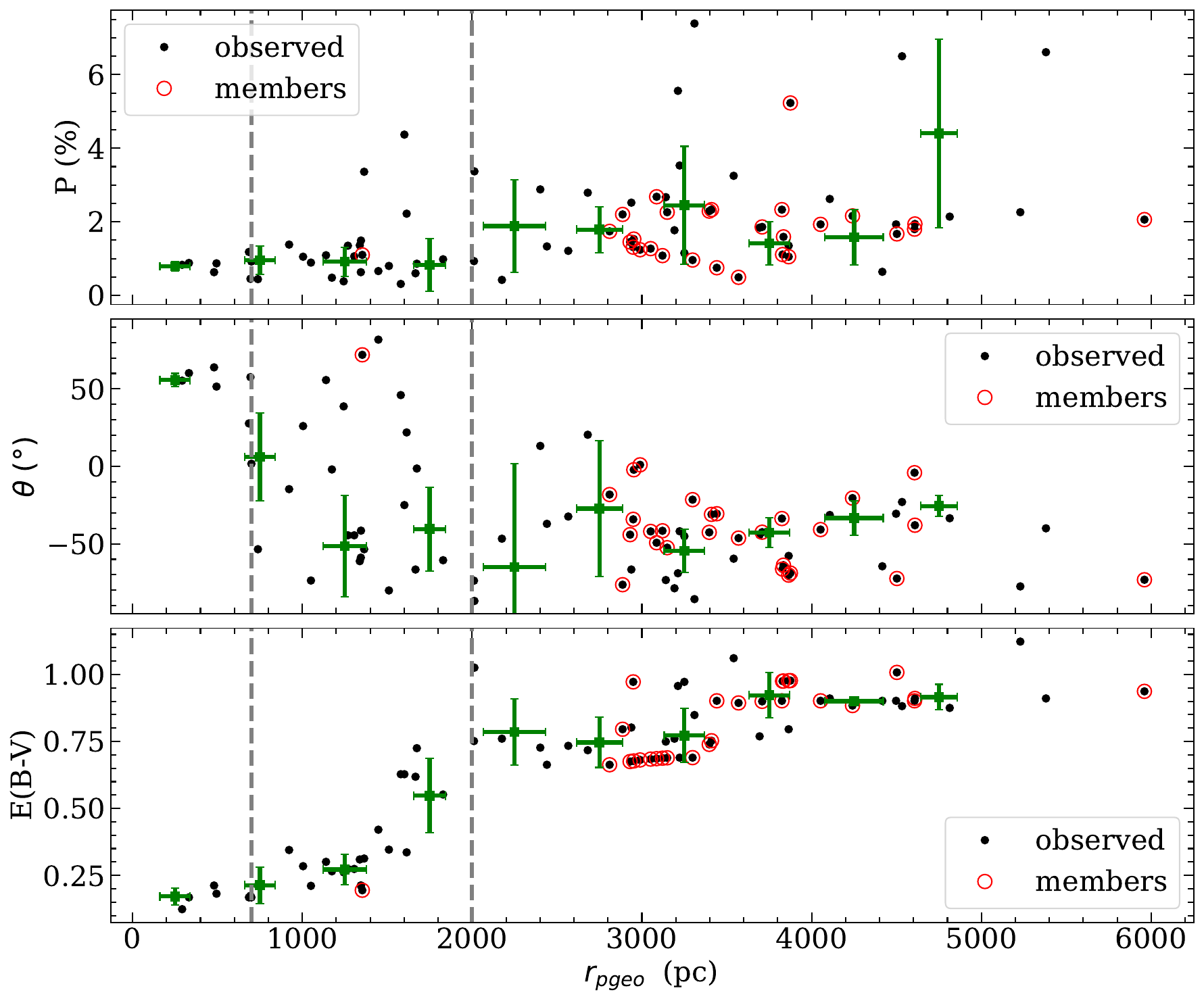}}
\subfigure[]
{\label{fig:5d}\includegraphics[width=0.44\textwidth]{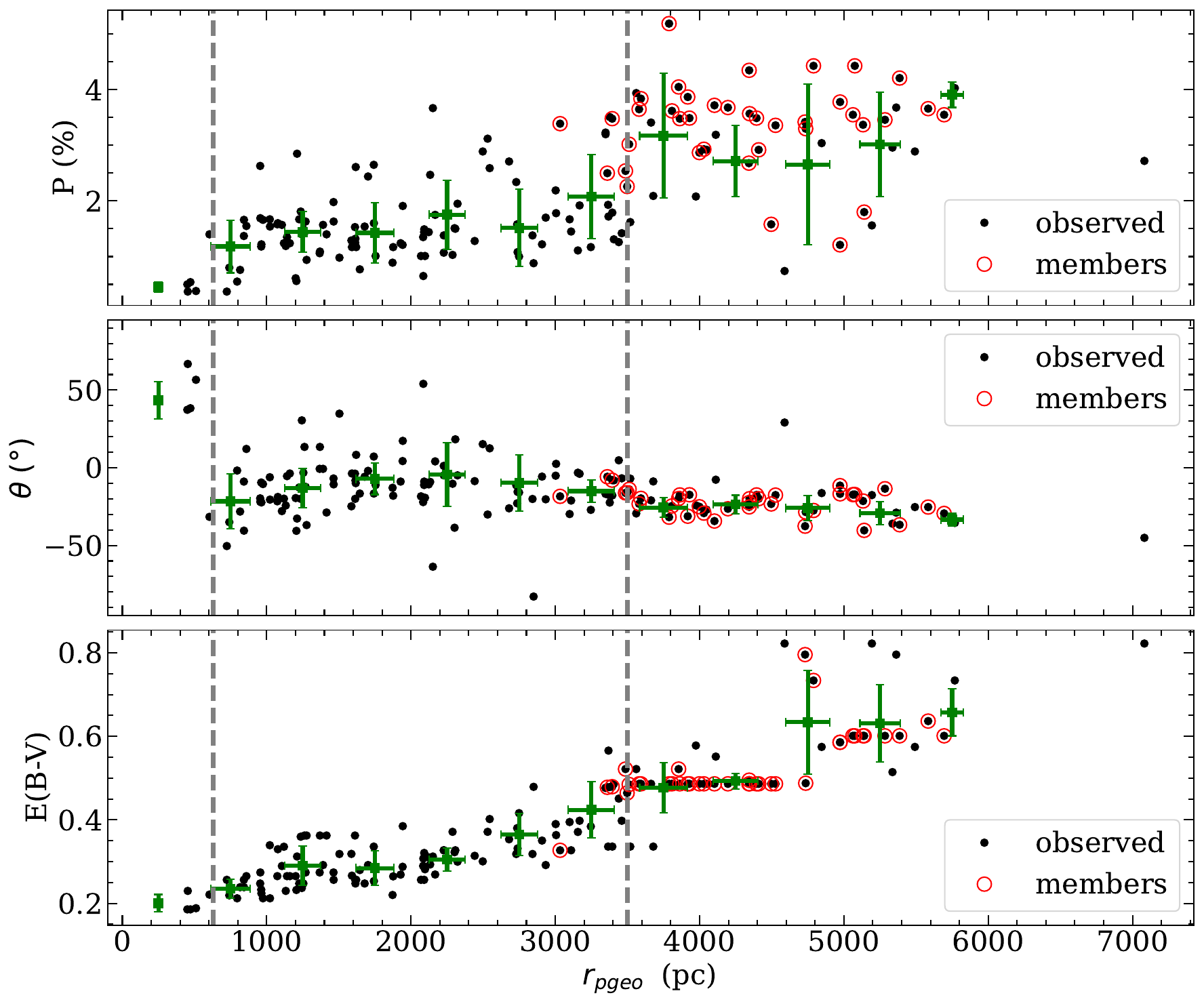}}
\subfigure[]
{\label{fig:5e}\includegraphics[width=0.44\textwidth]{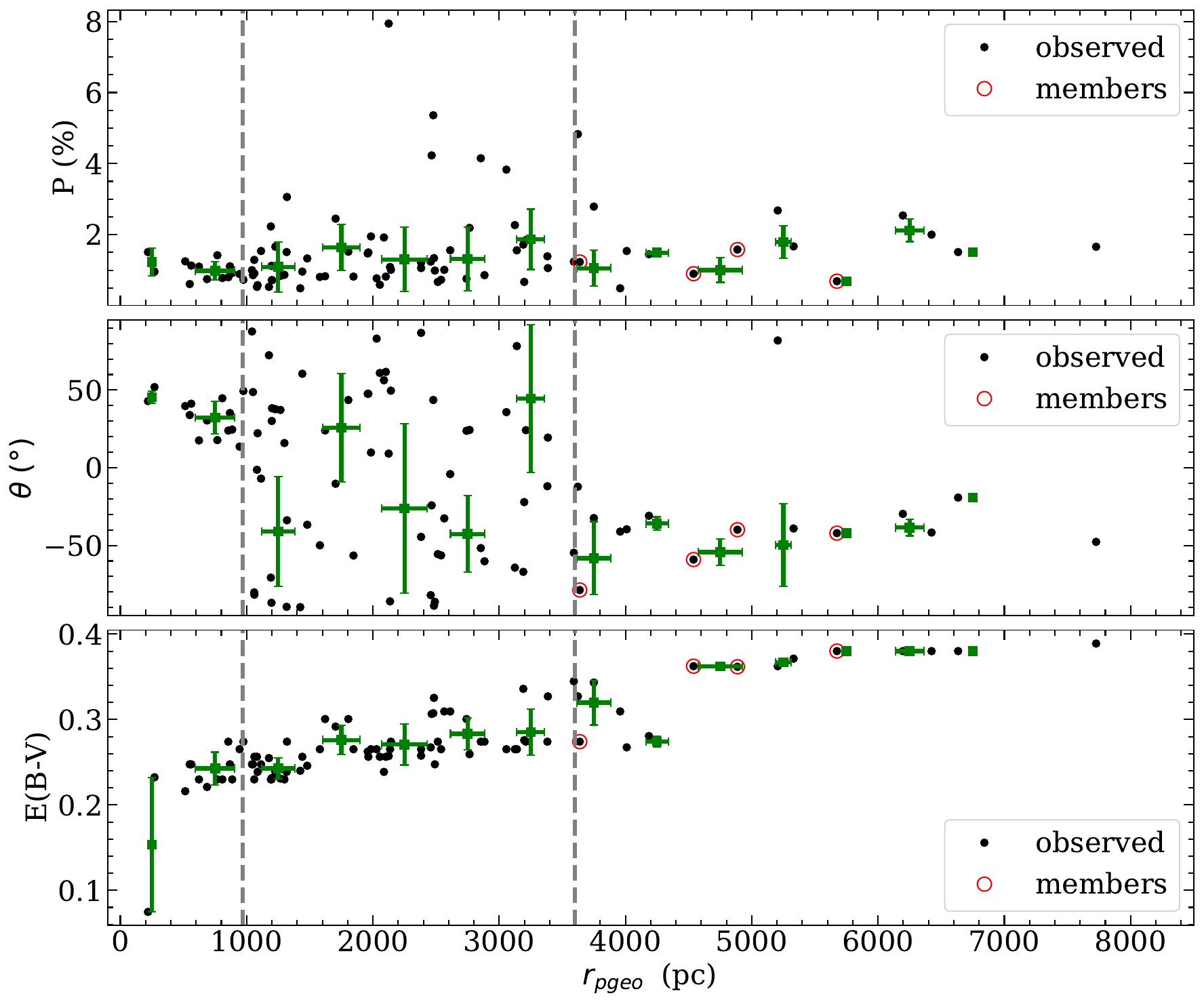}}
\caption{Variation of the degree of polarization ($P$), polarization angle ($\theta$), and reddening E(B-V) as a function of distance, $r_{pgeo}$ for clusters, panel (a) - Kronberger 1; panel (b) Berkeley~69; panel (c) Berkeley~71; panel (d) King~8; and panel (e) Berkeley~19 respectively. The vertical dashed lines in each panel denote the location of the foreground dust layers derived from visual inspection.} 
    \label{fig:Fig6}
\end{figure*}

To determine the dust distribution towards each cluster, we cross-matched the observed stars with the \textit{Gaia} EDR3 \citep{EDR3} distance catalog \citep{bailer2021} and used `$r_{pgeo}$' as the distance estimator of crossmatched sources. The extinction or reddening $E(B-V)$ information towards each star is extracted from Green's Bayestar 3D extinction map \citep{Green2019}. In Fig.~\ref{fig:Fig6}, we show the variation of the degree of polarization (top), polarization angle (middle), and reddenning $E(B-V)$ (bottom) as a function of `$r_{pgeo}$' distance towards the clusters Kronberger~1, Berkeley~69, Berkeley~71, King~8, and Berkeley~19 in panel~\ref{fig:5a} to \ref{fig:5e} respectively. 
Stars with membership probability greater than 50\% are shown using red open circles. We binned the data along the line of sight for each cluster direction in the distance axis to reveal the smooth variation in polarization and extinction with distance. The weighted average degree of polarization, circular weighted average polarization angle, and average extinction within 500 pc distance ($r_{pgeo}$) bins are depicted as green square points with error bars representing the weighted dispersion in both horizontal and vertical axes within the corresponding bin. 
 The individual point-wise error bars are omitted in the figures to prevent visual clutter and potential confusion. We refer to Figs.~\ref{fig:FigA1} to~\ref{fig:FigA5} in Appendix~\ref{sec:app} for the depiction of the degree of polarization and polarization angle for each cluster, along with their associated uncertainties in polarization measurements and estimated distances.

It is noted from the middle panel of Fig.~\ref{fig:5a} that the foreground stars towards the direction of Kronberger~1 exhibit a low degree of polarization with a polarization angle largely deviated from the Galactic plane.  The polarization angle becomes almost constant beyond $\approx$500 pc.
Despite the limited number of stars in this region, the abrupt change in the polarization angle implies the likely existence of a cloud layer around 500~pc (marked by a dotted vertical line in Fig.~\ref{fig:5a}). This transition aligns with observed jumps in extinction values, as evident in the bottom panel of Fig.~\ref{fig:5a}. Subsequently, polarization uniformly increases with distance until around 1600~pc (also marked by a vertical dotted line), where starlight undergoes systematic depolarization, leading to a reduction in the degree of polarization. This region coincides with the cluster position, as denoted by the location of members in the figure. The cluster is situated in close proximity to the star-forming region S234 \citep{Lokesh2018}. The decrease in the degree of polarization could be attributed to the cumulative effect of dust in the ISM and dust present in the star-forming region. However, the polarization angles remain relatively unchanged, possibly influenced by the dominant foreground dust layer present at around 500 pc.
An increase in the degree of polarization is observed beyond 3000~pc, which could indicate the presence of another dust layer at that distance. Similarly, multiple jumps in the degree of polarization, polarization angle, and extinction are evident towards each cluster direction (marked by dashed lines in Fig.~\ref{fig:5b} to  Fig.~\ref{fig:5e}), implying the existence of multiple dust layers along each line of sight. The second column of Table~\ref{tab:clouds} lists the approximate distance to the dust layers visually determined towards each cluster. The analysis reveals the presence of at least two dust layers in each cluster direction. While the potential presence of more than two dust layers exists, we focus on layers showing a clear change in the degree of polarization and extinction.
\begin{table}
    {\centering
    \caption{Distances of detected foreground dust clouds from visual inspection and \texttt{BISP-1}.}
    \label{tab:clouds}
    \begin{tabular}{lcc} % four columns, alignment for each
        \hline
        \hline \\[-1.5ex]
        Cluster & \multicolumn{2}{c}{Distance of inferred dust layers [pc]} \\ \\[-2.ex]
        & Visual inspection & \texttt{BISP-1} algorithm\tablefootmark{**} \\ \\[-2.0ex]
        \hline \\[-1.5ex]
    Kronberger~1  & 500, 1600, 3000 & 280, 686, 4035\\
    Berkeley~69  & 600, 2100 & 238, 511, 2183 \\
    Berkeley~71  &  700, 2000  &  245, 653, 2402 \\
    King~8 &  630, 3500  & 396, 756, 4396 \\
    Berkeley~19  & 970, 3600  &  193, 979, 3849\\ \\[-2.ex]
        \hline
    \end{tabular}
    \newline}
    \tablefoottext{**}{The reported distance to the first layer corresponds to the 95$^{th}$ percentile value while the median of the posterior is considered for the other layers.}
    %\textsuperscript{**}\footnotesize{The reported distance to the first layer corresponds to the 95$^{th}$ percentile value while the median of the posterior is considered for the other layers.}
    
\end{table}

\subsubsection{\texttt{BISP-1} method}\label{sec:5.2.2}
The visual examination of change in polarization and extinction of stars as a function of distance provides qualitative analyses, offering indications of the number of dust layers along a specific line of sight. 
However, accurately determining the distance to the dust layers and their associated uncertainties remains challenging.
Recently, \citet{Pelgrims2023} introduced a Bayesian method to decompose the starlight polarization along the distance. Their method, named \texttt{BISP-1} for Bayesian Inference of Starlight Polarization in 1D, is specifically designed to quantify the number of dust layers, their distance, and their polarization properties based on the linear polarization measurements of stars along with their distance data.
Unlike visual analyses, \texttt{BISP-1} accounts for statistical uncertainties associated with the estimated parameters derived from the posterior distribution of the dust layers.
This method was designed to work at high-galactic latitudes towards the diffuse ISM in the framework of the upcoming \textsc{Pasiphae} polarization survey \citep{Tassis2018}. \texttt{BISP-1} is the first Bayesian inference method developed for the tomographic decomposition of the plane-of-sky magnetic field component. The algorithm incorporates the polarization signal from the magnetized and dusty ISM, which is modeled by thin layers at different distances. It uses directly observed quantities such as Stokes $q$ and $u$ parameters, parallax, and their associated uncertainties as input data rather than relying on derived polarization and distance information. 

We employed the \texttt{BISP-1} algorithm on our polarization results to confirm and quantify the distance of the dust layers present along the line of sight. Here, we utilized the measured Stokes parameters $(q, u)$ of all the stars along with their parallax values from \textit{Gaia} DR3 \citep{GaiaDR3} without any constraints on the polarization S/N. However, we applied some quality criteria: % flags to ensure the reliability of a few parameters.
we considered stars with positive parallax values and a renormalized unit weight error (RUWE) less than 1.4. A RUWE value above 1.4 indicates the presence of blended sources, which may lead to unreliable parallax measurements. \texttt{BISP-1} assumes all data points to be associated with the dusty magnetized ISM, therefore we opted to eliminate outliers from the polarization measurements, potentially linked to intrinsic polarization. This was accomplished through a recursive sigma-clipping on the distribution of the Stokes parameters within groups of five neighbors defined in the physical 3D position space. Outliers were identified as stars for which the probability of their Stokes parameters originating from the same parent distribution as their neighbors fell below 1\%, accounting for distribution scatter and observation uncertainties. Upon inspecting the resulting selection for all line-of-sight samples, we observed that in the sample towards Kronberger~1, the star closest to the Sun, exhibiting polarization compatible with zero and isolated in distance space, was likely not an outlier. Hence, we have not considered that star as an outlier. We discarded the remaining outliers for subsequent analyses. 

We aimed at using \texttt{BISP-1} to test various models involving different numbers of clouds along the sightlines and, therefore, first needed to define our priors.
For this, we visualized the data in the $(q,\,u)$-$\mu$ space, where $\mu$ is the distance modulus defined as $\mu = 5\log_{10}(d) - 5$, with $d$ denoting the stellar distance in pc.
We noticed that for every sample, except Kronberger~1, the Stokes parameters of the nearest stars (first star according to distance) do not coincide with $(q,\,u) = (0,0)$, even when considering uncertainties.
This indicates the presence of a dust layer in the foreground of our samples (discussed below). In addition to this, we observed evidence for a change of polarization properties and extinction at a distance of 500-1000~pc (corresponding distance modulus $\mu \in [8,10] $) towards all the cluster directions (see Fig.~\ref{fig:Fig6}). Therefore, it became evident that each sightline intersected at least two dust clouds. Based on these observations, we defined our priors for the cloud parallaxes as follows: a flat prior for the parallax of a nearby layer with a minimum distance set to 50~pc and a Gaussian prior for the parallax of the second layer with a mean of 1.58~mas, corresponding to a distance modulus of $\mu = 9$, and a standard deviation of 0.5~mas. In addition, we used flat priors with the default setting in \texttt{BISP-1} for subsequent distant layers.

After defining the priors, we applied \texttt{BISP-1} to model the data using models from two to five dust layers. In each case, we ran a maximum of 40000 iterations with a required log-evidence precision of 0.01. The Akaike Information Criterion (AIC) was employed to assess the performance of the various models and identify the optimal one (see \citealt{Pelgrims2023, Pelgrims2024} for details).

\begin{figure*}
\centering
   \includegraphics[width=15cm]{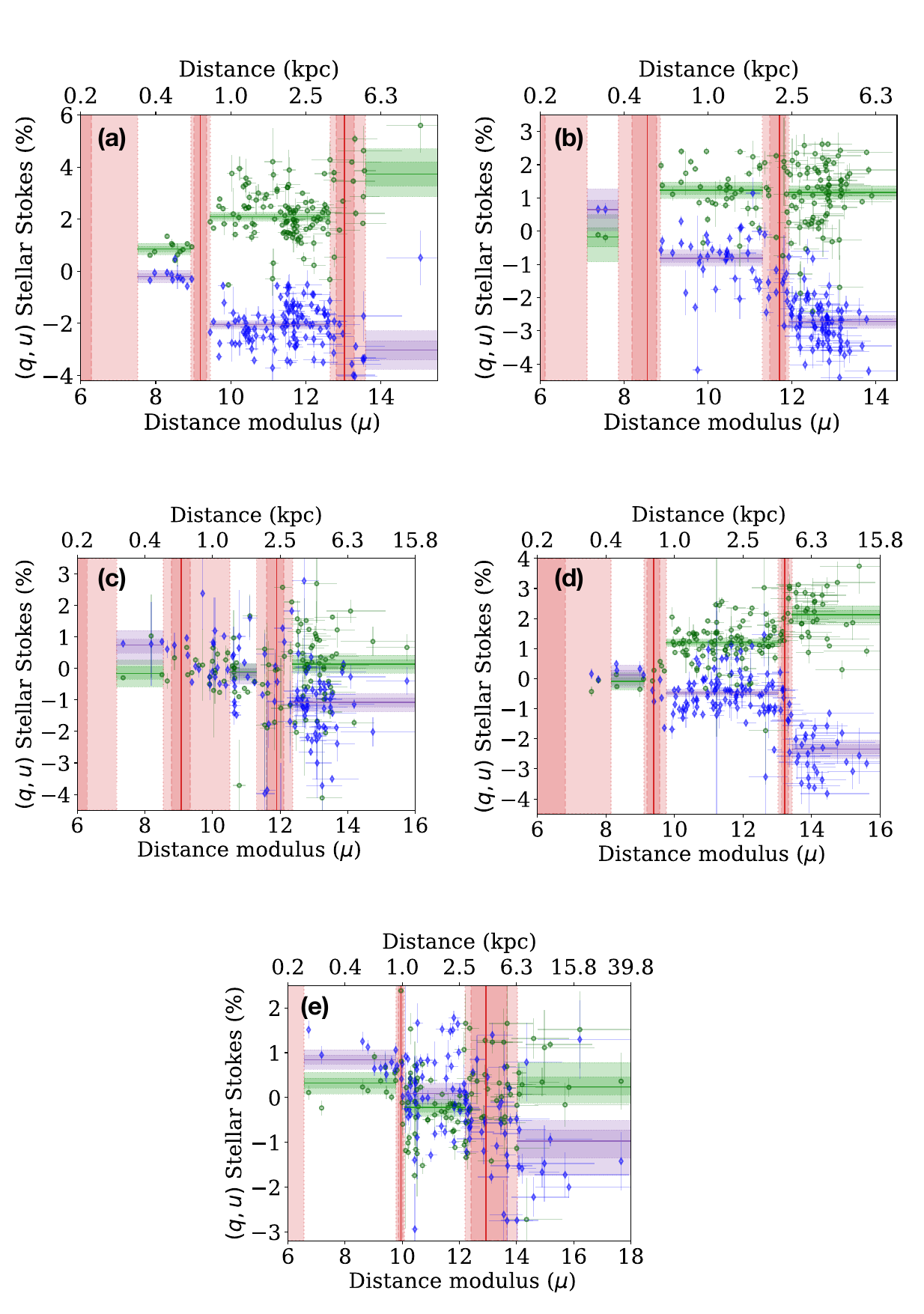}
    \caption{The distribution of $q$ (in green) and $u$ (in blue) parameter of the stars as a function of their distance modulus ($\mu$) or distance towards Kronberger~1 in (a), Berkeley~69 in (b), Berkeley~71 in (c), King~8 in (d), and Berkeley~19 in panel (e).
qThe red solid vertical line indicates the median of the posterior distribution on the cloud distance while the light and dark shade area denotes the range of $2.5$ to $97.5$ percentiles and $16$ to $84$ percentiles of the distribution. The green and purple horizontal lines indicate the median value of $q$ and $u$ from the posterior distribution of the mean Stokes parameters of the foreground dust clouds.  Shaded areas represent the $2.5$ to $97.5$ and $16$ to $84$ percentiles of the distributions.}  \label{fig:Fig7}
\end{figure*}

The results obtained from the \texttt{BISP-1} method for each cluster direction are presented in Fig.~\ref{fig:Fig7}. The figure illustrates the variation of $q$ (in green) and $u$ (in blue) as a function of distance modulus ($\mu$) or distance estimated from the inverse of parallax. The solid vertical lines in the figure denote the distance moduli corresponding to the median of the posterior distributions on dust-layer parallaxes. The 16th and 84th %$16^{th}$ and $84^{th}$ 
percentiles of the same distribution are marked by the dark-shaded region, while the $2.5$ and $97.5$ percentiles are denoted by the lightly shaded area.
The horizontal shaded areas show the (cumulative) posterior distributions of the mean Stokes parameters from the dust layers. The figure clearly demonstrates the presence of multiple dust layers in each cluster direction. 
The estimated distances to the line of sight dust clouds are listed in the third column of Table~\ref{tab:clouds} for comparison with the visual inspection. 
The degree of polarization and polarization angle derived from the resulting average $q$ and $u$ parameters, along with the predicted distance of each dust layer, are tabulated in Table~\ref{tab:BISPpol}.  

%\begin{figure*}
%\centering
%\subfigure[]
%{\label{fig:7a}\includegraphics[width=0.40\textwidth]{revised2/Fig6_R2_Kron1.pdf}}
%\subfigure[]
%{\label{fig:7b}\includegraphics[width=0.40\textwidth]{revised2/Fig6_R2_Be69.pdf}}
%\subfigure[]
%{\label{fig:7c}\includegraphics[width=0.40\textwidth]{revised2/Fig6_R2_Be71.pdf}}
%\subfigure[]
%{\label{fig:7d}\includegraphics[width=0.40\textwidth]{revised2/Fig6_R2_King8.pdf}}
%\subfigure[]
%{\label{fig:7e}\includegraphics[width=0.40\textwidth]{revised2/Fig6_R2_Be19.pdf}}
%\caption{The distribution of $q$ (in green) and $u$ (in blue) parameter of the stars as a function of their distance modulus ($\mu$) or distance towards Kronberger~1 in (a), Berkeley~69 in (b), Berkeley~71 in (c), King~8 in (d), and Berkeley~19 in panel (e).
%The red solid vertical line indicates the median of the posterior distribution on the cloud distance while the light and dark shade area denotes the range of $2.5$ to $97.5$ percentiles and $16$ to $84$ percentiles of the distribution. The green and purple horizontal lines indicate the median value of $q$ and $u$ from the posterior distribution of the mean Stokes parameters of the foreground dust clouds.  Shaded areas represent the $2.5$ to $97.5$ and $16$ to $84$ percentiles of the distributions.}

%    \label{fig:Fig7}
%\end{figure*}

In Fig.~\ref{fig:Fig7}, it is observed that a dust layer, including non-zero polarization to background stars, exists toward each cluster at a distance of $\mu \lesssim 6.5$ ($d \lesssim 200$~pc). This layer is not anticipated in visual inspection (Sect.~\ref{sec:5.2.1}) as we have very few stars with distances smaller than 400~pc in our samples. 
However, the posterior distribution of the parallax of the nearby cloud as estimated by \texttt{BISP-1} is not solely determined by the data at large parallax values (at low distances) but rather by a combination of the assumed prior and the nearest data points. These data points play a crucial role in determining the upper limit on the distance of the nearest clouds. Hence, for the nearest cloud layer, the upper limit on the distance estimate is considered and listed in Table~\ref{tab:clouds}. The lower limits are, however, strongly dependent on the choice of the upper limit of the parallax of this nearby cloud.

Moreover, the indication of nearby dust components towards each cluster is also compatible with the existence of the Local Bubble, the most prominent structure of the local ISM in which the Sun is embedded. The distance from the Sun to the inner shell of the dusty wall surrounding the cavity has been estimated in \citet{Pelgrims2020} from the 3D dust density map of \citet{Lallement2019}. Towards our sightlines, the dusty wall of the Local Bubble is estimated\footnote{In the sky area of interest, the inner shell of the Local Bubble is found to be at a distance of 175 to 195~pc from the Sun, determined from \url{https://doi.org/10.7910/DVN/RHPVNC} \citep{Data2022} with some added margin to account for the thickness of the Local Bubble wall.} to be in the range of $\sim$180~pc to 220~pc. The upper limits to the first cloud listed in column~3 of Table~\ref{tab:clouds} are consistent with this picture. 
\begin{table}
    {\centering
    \caption{Results of \texttt{BISP-1} characterizing degree of polarization, polarization angle, and distance of the dust clouds present along the line of sight of each cluster.}
    \label{tab:BISPpol}
    \begin{tabular}{lccc} % four columns, alignment for each
        \hline
        \hline \\[-1.5ex]
        Cluster & P $\pm$ eP  & $\theta$ $\pm$ e$\theta$ & median distance  \\
        & $(\%)$ & ($^\circ$) & (pc) \\
        \\[-2.0ex]
        \hline \\[-1.5ex]
        
    Kronberger~1  & 0.88  $\pm$ 0.12  & -6.6 $\pm$ 3.9 &  86  \\
    & 2.22 $\pm$ 0.06 &  -28.1  $\pm$  1.8 &  686\\
    & 1.91 $\pm$ 0.24 &  -15.4 $\pm$ 6.3 & 4035\\ \\[-1.7ex]
    Berkeley~69 &  0.67 $\pm$ 0.46  & 52.7 $\pm$ 14.3 & 85 \\
    & 2.03 $\pm$ 0.17 & -23.09 $\pm$ 4.8 &  511 \\
    & 1.91 $\pm$  0.07 &  -45.0 $\pm$ 2.5 &  2183\\  \\[-1.7ex]
    Berkeley~71  &   0.74 $\pm$ 0.33 & 50.9 $\pm$ 8.2 & 94 \\
    & 0.87 $\pm$  0.31 & -44.0 $\pm$ 8.2 & 653 \\
    & 0.96 $\pm$  0.22 & -37.19 $\pm$ 6.5 & 2402\\  \\[-1.7ex]
    King~8 &  0.13 $\pm$ 1.25 & 66.0 $\pm$ 35.0  & 92 \\
    & 1.40 $\pm$ 0.11 &  -12.2 $\pm$  3.9 & 756 \\
    &  2.09 $\pm$ 0.08 & -31.6 $\pm$ 2.2 & 4396\\ \\[-1.7ex]
   Berkeley~19  &  0.91 $\pm$ 0.12 & 34.4 $\pm$ 3.7 & 85 \\ 
    &  0.94 $\pm$ 0.17 & -62.5 $\pm$  4.7 & 979\\
    &  1.21 $\pm$ 0.23 & -33.3 $\pm$ 5.7 &  3849\\   
    \\[-2.ex]
        \hline
        
    \end{tabular}
    }
\end{table}

\subsection{Signature of spiral arms}\label{sec:5.3}
\begin{figure}%[htb!]
\centering
	\includegraphics[scale = 0.35]{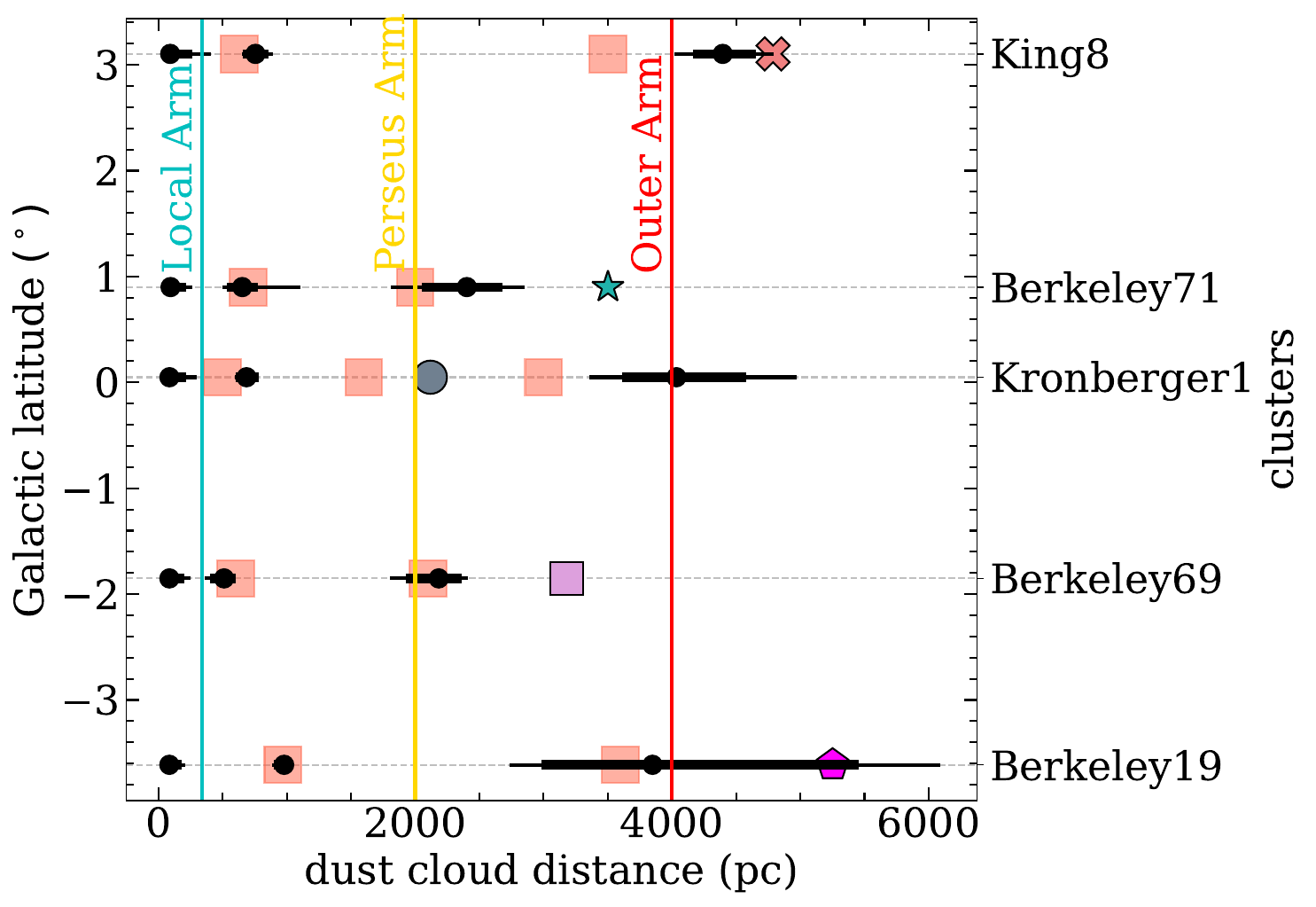}
    \caption{The distance of the dust layers along the line of sight of observed cluster directions inferred from visual inspection (in the light-red squares) and \texttt{BISP-1} algorithm (as black points). The thick black line around the black point corresponds to the 16th and 84th 
    percentiles of the cloud distance while the thin line represents the 2.5 and 97.5 percentile ranges. The cluster distances are marked by different color symbols at their respective Galactic latitude positions. The vertical lines represent the location of the Spiral arms: the Local arm (cyan), the Perseus arm (yellow), and the Outer arm (red). }
    \label{fig:Fig8}
\end{figure}

The analyses presented in Sect.~\ref{sec:5.2} reveal the presence of multiple dust layers towards each cluster direction, regardless of the method employed. The distance of the predicted dust layers from different methods is cataloged in Table~\ref{tab:clouds}. The results are also shown in Fig.~\ref{fig:Fig8}, where we plotted the estimated distance of all dust layers resulting from visual inspection (in the light-red squares) and \texttt{BISP-1} algorithm (in black points). The error bar around the \texttt{BISP-1} inferred distance represents the 2.5 and 97.5 percentile (thin black line) and  16 and 84 percentile values (thick black lines). The solid vertical lines represent the approximate distance of the spiral arms towards the anticenter direction (Local arm: cyan, Perseus arm: yellow, and Outer arm: red) determined from \citet{Reid2019} and \citet{Ginard2021}. The cluster names are placed based on their relative Galactic latitudes, ranging from $-3^\circ.6$ to $3^\circ.1$, and their distances are marked in the figure using different color symbols. It is worth noting that \texttt{BISP-1} estimates the parallax and normalized Stokes parameters of dust layers along the line of sight. Consequently, the inverse of the parallax is used to estimate the distance of those clouds. However, we utilized $r_{pgeo}$ as the distance estimator in the visual inspection. For the distant stars (e.g., stars towards the King~8 and Berkeley~19 clusters), the uncertainty in the parallax measurement can increase, and the $r_{pgeo}$ distance will largely depend on the prior assumptions \citep{bailer2015estimating, bailer2021}.
Hence, the distant clouds may show large deviations in their distance estimations from visual inspection and \texttt{BISP-1} merely due to the use of $r_{pgeo}$ in one case and the inverse of parallax in the other case.
A careful inspection of Fig.~\ref{fig:Fig8} reveals that Berkeley~69 and Berkeley~71, which are located between the Perseus arm and the Outer arm, show the presence of a dust cloud at $\approx$~2~kpc. Interestingly, this location aligns with the anticipated position of the Perseus arm (depicted by the yellow line) in the anticenter direction. 
Additionally,  NGC~1893 ($\ell = 173^\circ.582$, $b = -1^\circ.659$, and r = 2.8~kpc), another cluster in a similar direction, has been reported to have a dust cloud at a distance of $\approx$~2~kpc \citep{Bijas2022}. This implies that the dust present in the Perseus arm is responsible for the polarization of background stars towards this direction. The Kronberger~1 cluster itself resides in the dusty environment of this arm, making it challenging to predict the signature of the arm towards this cluster.

Furthermore, the distant clusters, King~8 and Berkeley~19, located beyond the Outer arm, despite having opposite latitudes ($-3.6^\circ$ and $3.1^\circ$, respectively), both show a dust layer at approximately $3500-4500$~pc. The Outer arm of the Galaxy may also be located at $\approx$~4000~pc (solid red line, as predicted from the \citet{Reid2019} model on High Mass Star Forming Region, HMSFR data). This suggests that the major changes in polarization are occurring near the spiral arms of the Galaxy, highlighting polarization as an effective tool for understanding large-scale dust distribution and mapping the underlying structure of the Galaxy.

In addition, to our surprise, we noted that the polarization measurements and extinction values of stars towards  Berkeley~19 and King~8 do not exhibit strong signatures of the Perseus arm, contrary to our initial expectations. The absence of a prominent dust layer around the Perseus arm for these higher latitude clusters suggests potential variations in the thickness of the spiral arms. It could be possible that the vertical thickness of the Perseus arm might be smaller than that of the Outer arm, leading to less pronounced effects at higher latitudes.

This speculation is further supported by the fact that the Galactic disk exhibits a flare, i.e., the scale height of the Galactic disk increases with the Galactocentric distance (e.g., \citealt{Drimmel2001}; \citealt{Uppal_2023}). Therefore, the Outer arm is expected to have a greater thickness than the Perseus arm. 
According to the linear model derived from the flare observed in COBE-DIRBE dust maps by \citet{Drimmel2001}, 
the scale height of the Galactic disk is estimated to be $\approx$~220~pc at the location of the Perseus arm and $\approx$~260~pc at the Outer arm. However, the height of the disk ($Z = d\,\sin b$; where $d$ is the distance from the Sun and $b$ is the Galactic latitude) in the line of sight of King~8 ($b = 3.104^\circ$) and Berkeley~19 ($b = -3.612^\circ$) within the Perseus arm is calculated to be approximately $110-125$~pc and that of the Outer arm around $220-250$~pc. These values of the disk height are notably smaller than the computed scale height of the disk at the corresponding arms. 
This implies that the absence of polarization features around the Perseus arm at higher latitude sightlines may not be directly related to the global features, such as the flaring of the disk.  Alternatively, it could indicate that the flared geometry of the disk, as traced by dust, differs from the linear predictions made by \citet{Drimmel2001}.

Another plausible explanation could be the existence of a low extinction window within the Perseus arm, specifically towards King~8 and Berkeley~19.
The mid-IR images obtained from the James Webb Space Telescope provide evidence of variations in the dust distribution within the spiral arms for external galaxies \citep[e.g.,][]{Meidt2023}. It is possible that the sightlines towards our higher latitude clusters may have less dust in the Perseus arm to polarize the background stars. 
To investigate this possibility further, we require polarization observations of distant clusters beyond the Outer arm with large spatial coverage.
In any case, our polarization results provide insight into the signature of the large-scale and small-scale structures of the Galactic disk.

\begin{figure}
\centering
\subfigure{\includegraphics[width=0.42\textwidth]{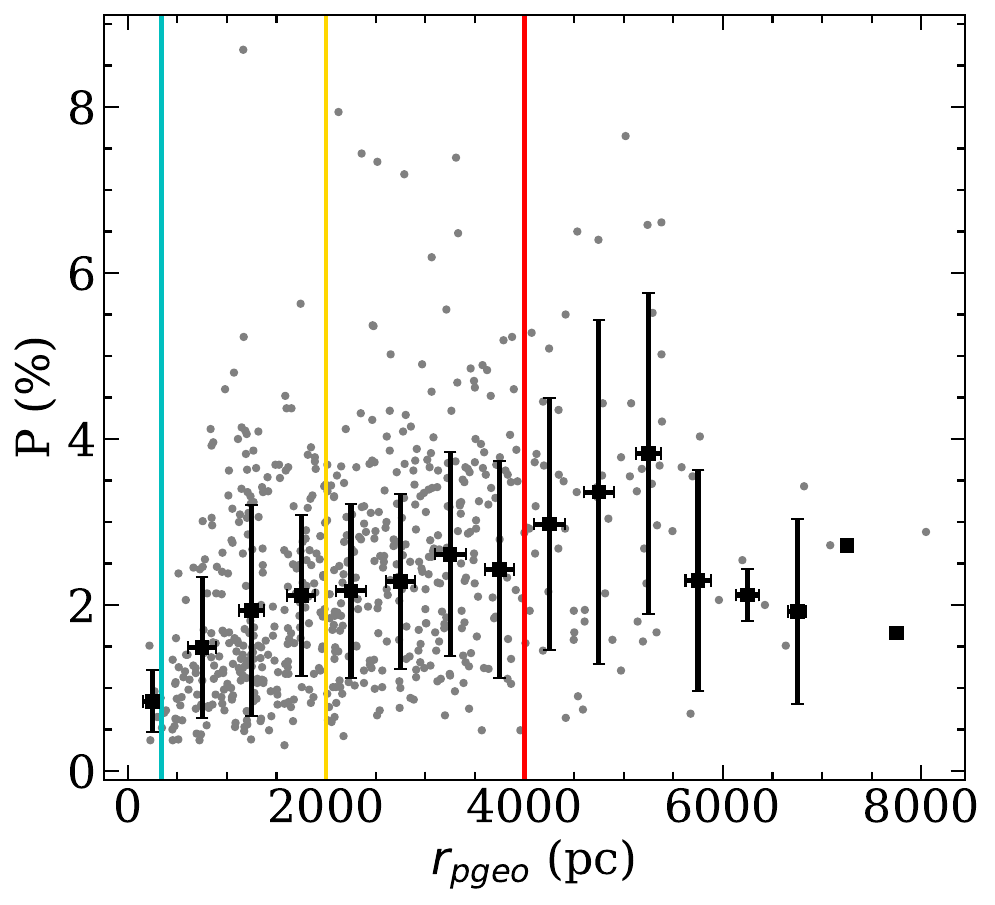}}
\subfigure{\includegraphics[width=0.42\textwidth]{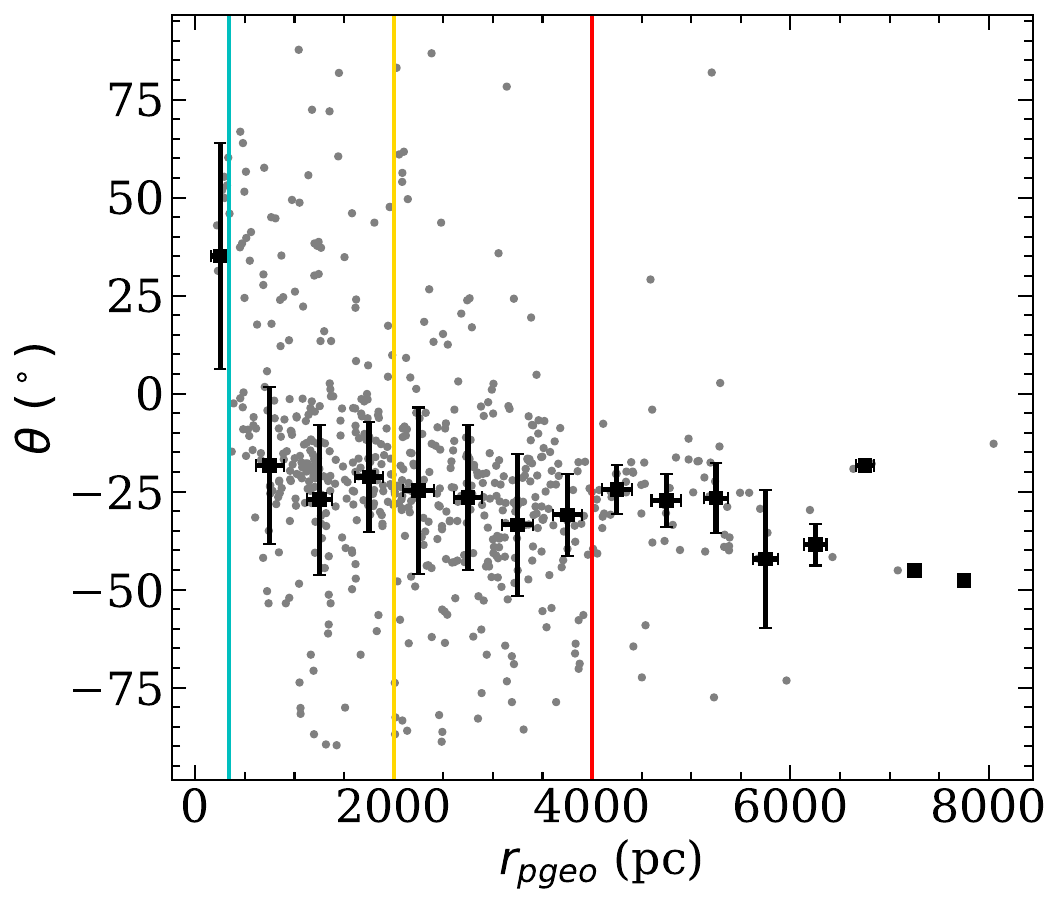}}
 \caption{Variation of degree of polarization (top panel) and polarization angle (bottom panel) with $r_{pgeo}$ for all the stars (members as well as non-members), observed towards the anticenter direction. The black points with error bars represent the average $P$ and $\theta$ values in $r_{pgeo}$-distance bins of 500~pc width. The error bars correspond to the standard deviation in the $P$ and $\theta$ values in each distance bin. The distance to the spiral arms is marked by vertical lines using the same colors as in Fig.~\ref{fig:Fig8}.}
    \label{fig:Fig9}
\end{figure}

In addition, all the polarization data observed towards the anticenter direction in this study is plotted in Fig.~\ref{fig:Fig9} displaying the variation of the degree of polarization (in the upper panel) and polarization angle (in the lower panel) with distance ($r_{pgeo}$). 
 The figure encompasses the data from a broad sky area ($\sim 7^\circ$ diameter) and shows a large scatter ($-90^\circ \, {\rm{to}} \, +90^\circ$) in the polarization angle. To see the variation of polarization with distance, we calculated the circular weighted average polarization angle and weighted mean degree of polarization in 500 pc bins of distance, $r_{pgeo}$.
 The weighted means are represented by the black squares with the error bars depicting the standard deviation of polarization values in each bin.  The mean polarization ($P$ as well as $\theta$) values exhibit slight changes in proximity to the spiral arms (vertical lines with color-coding consistent with Fig.~\ref{fig:Fig8}), thus highlighting the presence of spiral arms in the polarization data.

\subsection{General trend of dust distribution}
\label{sec:5.5}
\begin{table*}
\centering
\caption{The parameters of ten Galactic clusters towards anticenter direction, having polarization measurements from our observations and the literature. }\label{tab:3}
\begin{tabular}{lrrrrrrrcccl}
\hline
\hline \\[-1.5ex]

  \multicolumn{1}{c}{cluster} &
  \multicolumn{1}{c}{$\ell$} &
  \multicolumn{1}{c}{$b$} &
  \multicolumn{1}{c}{dis} &
  \multicolumn{1}{c}{$P_{av}$} &
  \multicolumn{1}{c}{$\sigma_{P_{av}}$} &
  \multicolumn{1}{c}{$\theta_{av}$} &
  \multicolumn{1}{c}{$\sigma_{\theta_{av}}$} &
  \multicolumn{1}{c}{members} & 
  \multicolumn{1}{l}{Reference for polarization} \\

  & (deg) & (deg) & (kpc) & ($\%$) & ($\%$) & (deg) & (deg) & & measurements\\ \\[-1.7ex]
 \hline \\[-2.ex]
  Kronberger~1 & 173.106 & 0.049 & 2.12 & 2.55 & 0.61  & -27.2 & 4.9  & 19 & this work\\
  Berkeley~69 & 174.442 & -1.851 & 3.18 & 3.23 & 0.70  & -33.8 & 8.7 & 48 & this work\\
  Berkeley~71 & 176.635 & 0.901 & 3.50 & 1.50 & 0.53  & - 40.0 & 15.5 & 28 & this work\\
  King~8 & 176.384 & 3.104 & 4.79 & 3.13 & 0.99  & -25.8 & 7.2  & 38  & this work\\
  Berkeley~19 & 176.919 & -3.612 & 5.25 & 1.08 & 0.27  & -62.5 & 14.9  & 04  & this work\\[1.5ex] 
  NGC~2281 & 174.892 & 16.889 & 0.51 & 0.85 & 0.14 & 16.6 & 4.8 & 09 & 1   \\
  NGC~1960 & 174.544& 1.076 & 1.12 & 1.23 & 0.12 & -21.3 & 4.7  & 08 & 1 \\
  Stock~8 & 173.375& -0.190 & 1.96 & 2.60 & 0.53  & -25.1 & 7.0  & 05 & 1 \\
  NGC~1931 & 173.914 & 0.269 & 2.11 & 1.85 & 1.39 & -25.4 & 9.4  & 18 & 2\\
 NGC1~1893 & 173.582 & -1.659 & 2.85 & 2.70 & 0.33 & -21.9 & 4.7 & 22 & 3 \\
  \\[-1.5ex]
\hline
\end{tabular}
\tablefoot{The cluster parameters include cluster distance (dis), average degree of polarization ($P_{av}$), and polarization angle ($\theta_{av}$), calculated as the mean values of the corresponding parameters for member stars. $\sigma$ and $\epsilon$ represent the dispersion and propagated uncertainty in the polarization measurements for the member stars.}
\tablebib{
(1)~\citet{Eswaraiah2011}; (2) \citet{Pandey2013}; (3) \cite{Bijas2022}.
}
\end{table*}

In this section, we examine the large-scale dust distribution by combining our observed cluster data with the polarization data from other clusters located in a similar line of sight available in the literature. We incorporate the data of five clusters: NGC~2281, Stock~8, NGC~1960, NGC~1931, and NGC~1893, obtained from (\citealt{Eswaraiah2011}; \citealt{Pandey2013}; \citealt{Bijas2022}). Among these selected clusters, NGC~1931 and Stock~8 are located in close proximity to the star-forming region in the Perseus arm. NGC~2281 is a foreground cluster situated at a high latitude ($b = 16^\circ.9$) and a distance of approximately 500~pc. NGC~1960 is also a foreground cluster with a distance of 1.12~kpc. Only one cluster, NGC~1893, is situated between the Perseus and Outer arms.
We cross-matched these clusters with the \cite{Hunt2023} catalog to find the member stars having a membership probability of more than 50\%. The Galactic position of the clusters $(\ell, b)$, and their distance (dis) values obtained from \citet{Hunt2023} are tabulated in the second to fourth columns of Table~\ref{tab:3}. The number of member stars detected (col: members) after cross-matching, their weighted average degree of polarization ($P_{av}$), and circular weighted mean polarization angle ($\theta_{av}$) in equatorial coordinates are also listed in the table, along with their weighted standard deviation ($\sigma_{P_{av}}$ and $\sigma_{\theta_{av}}$). Thus, we get the polarization information of a total of ten Galactic open clusters (five from the literature and five from our observation) in the anticenter direction, distributed over a large distance range.
\begin{figure}
\centering
\subfigure{\includegraphics[width=0.4\textwidth]{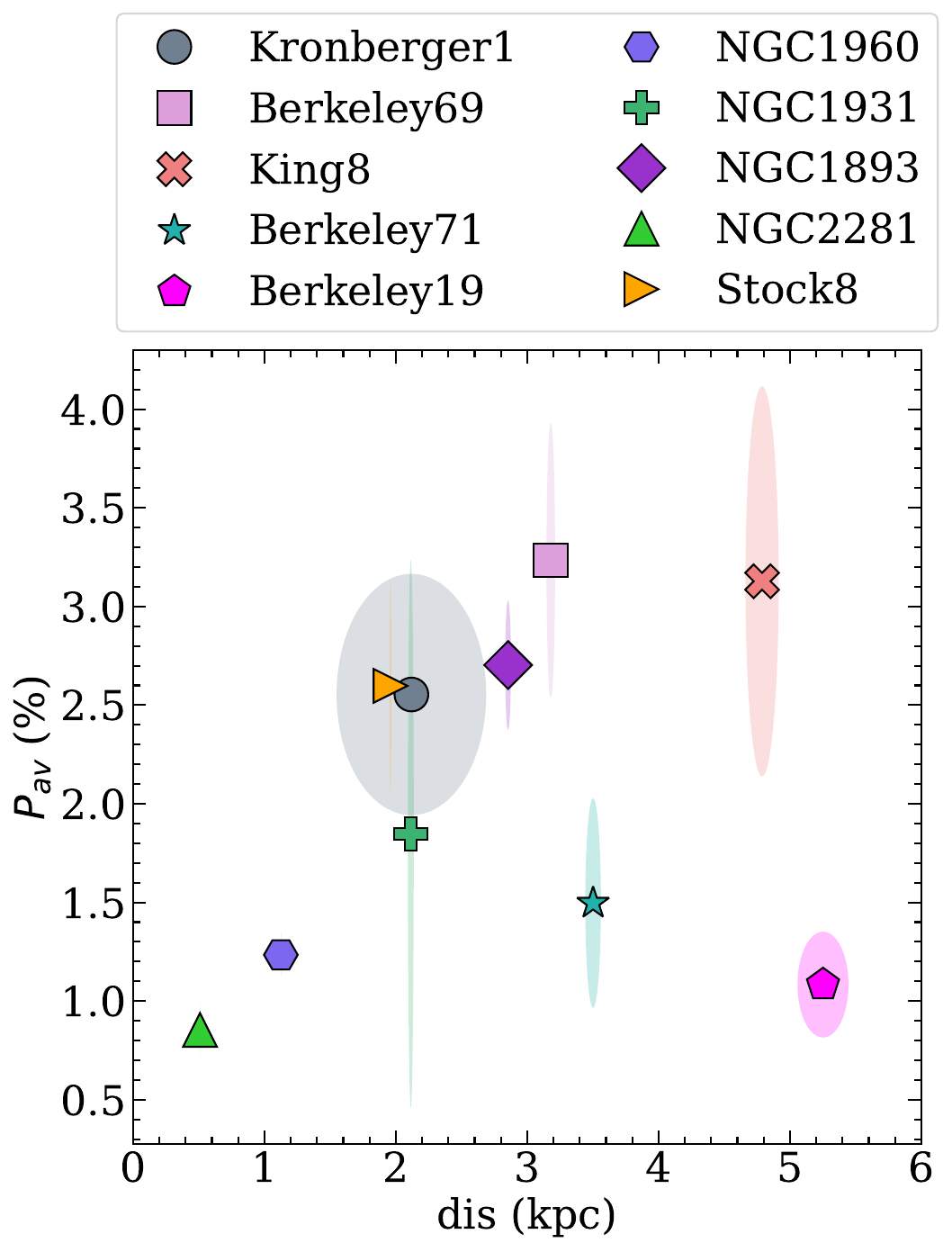}}
\subfigure{\includegraphics[width=0.4\textwidth]{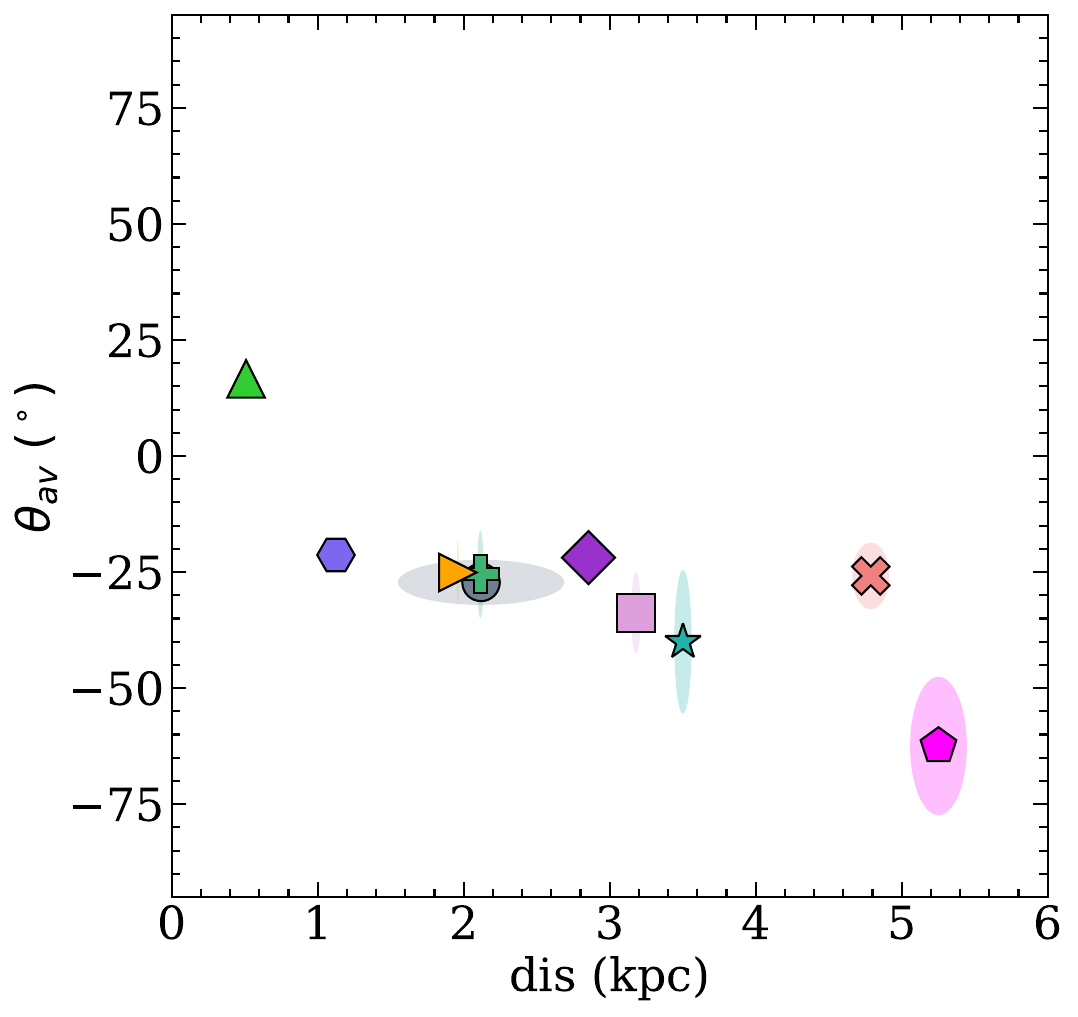}}
    \caption{Weighted average degree of polarization (top panel) and polarization angle (bottom panel) versus the distance of each cluster in different color symbols. The shaded ellipse corresponds to the weighted standard deviation in the corresponding axes.}
    \label{fig:Fig10}
\end{figure}
To investigate the variation of polarization of clusters with the distance, 
we plotted the weighted mean of the degree of polarization ($P_{av}$) and weighted circular mean of polarization angle ($\theta_{av}$) of cluster members with the average distance (dis) in the upper and lower panels of Fig.~\ref{fig:Fig10}. The different color symbols correspond to the different clusters. In these figures, the ellipse surrounding the central symbol represents the weighted standard deviation in polarization ($\sigma_{P_{av}}$ and $\sigma_{\theta_{av}}$, along the $y$-axis) of the member stars and the distance to the cluster \citep[determined from the upper bound given in][]{Hunt2023}. Figure~\ref{fig:Fig10} shows that the degree of polarization increases with distance while the polarization angle remains relatively constant except for NGC 2281, which is a high latitude cluster ($b = 16.887^\circ$) and Berkeley~19, where the average values are obtained based only on the four probable member stars. The uniform alignment of the polarization angle suggests that the plane-of-sky orientation of the magnetic field in the ISM towards this direction remains relatively constant while more dust column is traversed.
In this analysis, Berkeley~71 stands out as a special case where there is a localized variation in dust distribution in front of the cluster (as discussed in Sect.~\ref{sec:5.1}), resulting in a lower degree of polarization and a wider spread in the polarization distribution possibly due to depolarization and small-scale spatial variations in a foreground cloud (as seen in Fig.~\ref{fig:Fig4}b). Overall, the anticenter direction represents a low extinction direction with a relatively uniform magnetic field orientation parallel to the Galactic plane.

\section{Summary \& conclusions}\label{sec:6}
In this study, we conducted R-band polarization observations towards five clusters —Kronberger~1, Berkeley~69, Berkeley~71, Berkeley~19, and King~8— located in the anticenter direction ($\ell \approx 180^\circ$) covering a distance range of 2-6~kpc. These clusters were strategically chosen to infer the presence of dust layers towards the anticenter direction. Deciphering the distance of these layers along the line of sight becomes challenging when relying solely on radial velocity information derived from dust clouds. Additionally, the polarization measurements of the selected clusters contribute to our understanding of both small-scale and large-scale dust distribution along the line of sight and on the plane of the sky.

The polarization observations were carried out in the dark nights of October 2022 from AIMPOL instrument mounted on the 1.04~m Sampurnanand telescope of ARIES, Nainital. We obtain R-band polarization measurements of 159, 152,  80, 161, and 88 stars towards Kronberger~1, Berkeley~69, Berkeley~71, King~8, and Berkeley~19, respectively. Our analysis of the spatial dust distribution reveals the patchy distribution of warm dust in the sky region towards most of the clusters, as traced by the WISE W4 band. The deviation of polarization values of some of the member stars can be attributed to the presence of warm dust. 
Polarization measurements, together with the Herschel 250 $\mu$m wavelength intensity, trace the cold dust distribution towards Kronberger~1 and Berkeley~71.  The magnetic field orientation of the Berkeley~71 cluster helps to trace the morphology of the foreground dust cloud observed at Herschel 250 $\mu$m wavelength.

Furthermore, we analyzed the polarization measurements together with information on the distance to the observed stars using both qualitative and quantitative methods 
to disentangle the number of dust clouds along the line of sight. 
Our analysis reveals the existence of multiple (two or more) foreground dust layers toward all observed cluster directions. Distances of these layers were approximated from the jump in the degree of polarization, polarization angle, and extinction as a function of distance to the individual stars towards the corresponding clusters. Berkeley~69 and Berkeley~71 reveal the presence of a dust layer at approximately 2~kpc, coinciding in the distance with the Perseus arm. On the other hand, Berkeley~19 and King~8 demonstrate the existence of distant dust layers, likely associated with the Outer arm in the anticenter direction. The absence of a discernible Perseus arm signature in the higher latitude regions ($|b| > 3^\circ)$) suggests that the Perseus arm could be thinner than the Outer arm of our Galaxy. However, it could also be related to local structures within the Perseus arm. These results highlight the dust distribution over small as well as large-scale structures of the Galaxy.

Compilation of the polarization measurements in the anticenter direction, both from this study and the literature, reveals a consistent rise in the degree of polarization and a gradual stabilization of polarization angle with increasing distance, spanning approximately a $4^\circ$ radius. This uniform trend suggests a homogeneous dust distribution along this line of sight, characterized by a consistent alignment of the dust grains and the plane-of-sky component of the magnetic field.
Our findings indicate that the anticenter direction represents a region of low extinction, with low dust content, exhibiting minimal local variation. 
The observed changes in polarization and extinction arise primarily due to the presence of global features like the Galactic spiral arms. 
Our results strongly justify the need for a large-scale polarization survey to be carried out to trace the structure in the entire Galactic plane.

%\section*{Acknowledgements}
\begin{acknowledgements}
We thank the anonymous referee for their valuable and detailed reports that have significantly improved the quality of the manuscript. 
We thank Prof. Tassis, Department of Physics, University of Crete, for the helpful discussions.
We acknowledge the Telescope Time Allocation Committee of Sampurnanand Telescope (ARIES) for approving and allocating time for our proposal. We also thank the local staff and colleagues at the Sampurnanand telescope, ARIES, for their help during observations. Work at the Physical Research Laboratory is supported by the Department of Space, Govt. of India. 
We acknowledge useful discussions made possible by the Belgo-Indian Network for Astronomy (BINA) supported by the International Division, Department
of Science and Technology (DST, Govt. of India; DST/INT/BELG/P-09/2017)
and the Belgian Federal Science Policy Office (BELSPO, Govt. of Belgium;
BL/33/IN12).  

A part of this work has made use of data from the European Space Agency (ESA) mission \textit{Gaia}\footnote{https://www.cosmos.esa.int/web/gaia}, processed by the Gaia Data Processing and Analysis Consortium (DPAC\footnote{https://www.cosmos.esa.int/web/gaia/dpac/consortium}). Funding for the DPAC is provided by national institutions, in particular, the institutions participating in the Gaia Multilateral Agreement.
VP acknowledges funding from a Marie Curie Action of the European Union (grant agreement No. 101107047).  
This research has also made use of the NASA/IPAC Infrared Science Archive, which is funded by the National Aeronautics and Space Administration and operated by the California Institute of Technology. Astropy,\footnote{http://www.astropy.org} a community-developed core Python package for Astronomy \citep{astropyI, astropyII} was employed in the research. We have also used the VizieR catalog access tool, CDS, Strasbourg, France. 
\end{acknowledgements}
\bibliographystyle{aa} % style aa.bst
\bibliography{ref} % your references Yourfile.bib
%%%%%%%%%%%%%%%%%%%%%%%%%%%%%%%%%%%%%%%%%%%%%%%%%%
\begin{appendix}
\section{Variation of degree of polarization and polarization angle with distance}\label{sec:app}
\begin{figure*}
\centering
\subfigure[]{\label{fig:A1a}\includegraphics[width=0.44\textwidth]{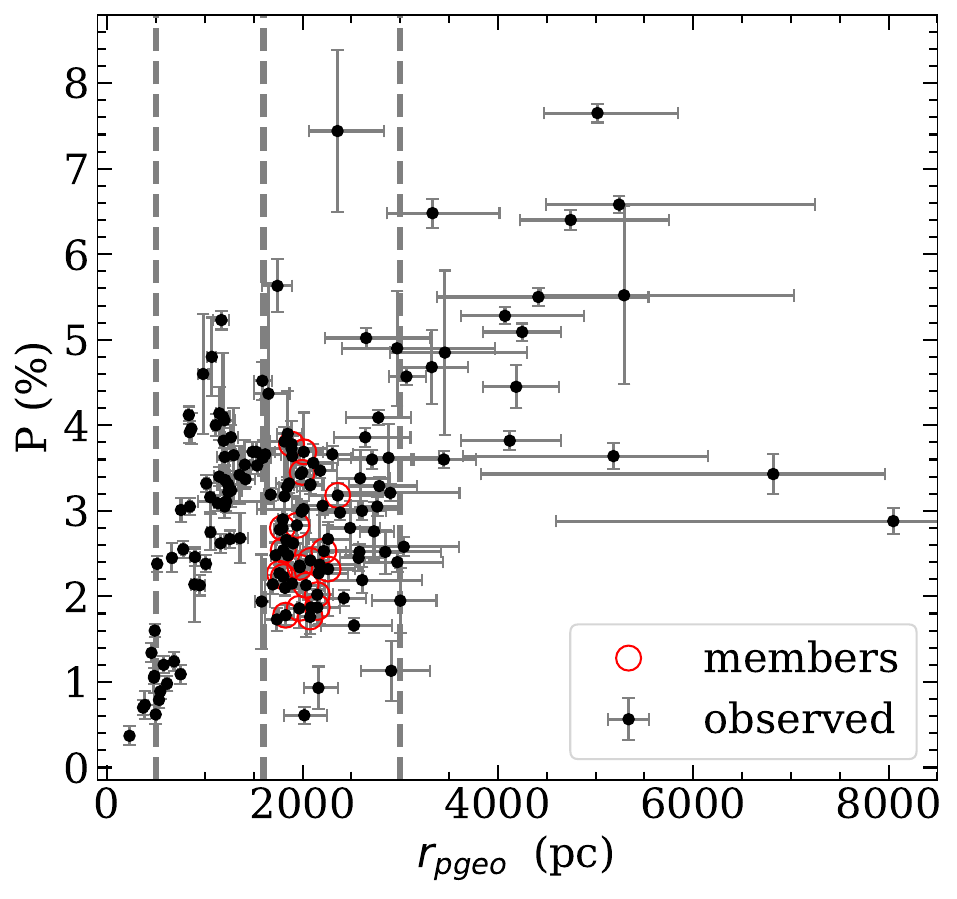}}
\subfigure[]{\label{fig:A1b}\includegraphics[width=0.47\textwidth]{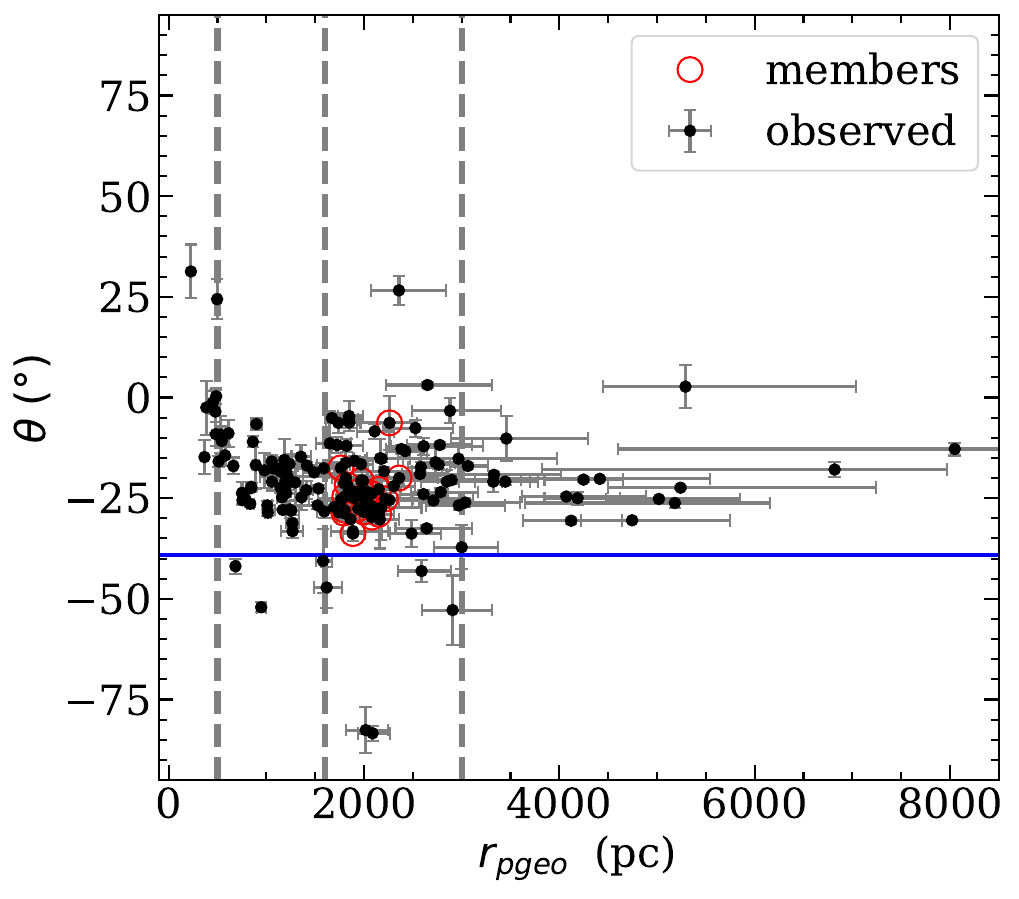}}
 \caption{Variation of the degree of polarization ($P$, a), and polarization angle ($\theta$, b) as a function of distance, $r_{pgeo}$ for Kronberger~1 cluster direction. The probable cluster member stars are marked by the red open circles. The solid blue line in panel (b) represents the orientation of the Galactic plane in this direction, and the vertical gray lines in both panels correspond to the expected distance to the foreground clouds. }\label{fig:FigA1}
\end{figure*}

\begin{figure*}
\centering
\subfigure[]{\label{fig:A2a}\includegraphics[width=0.44\textwidth]{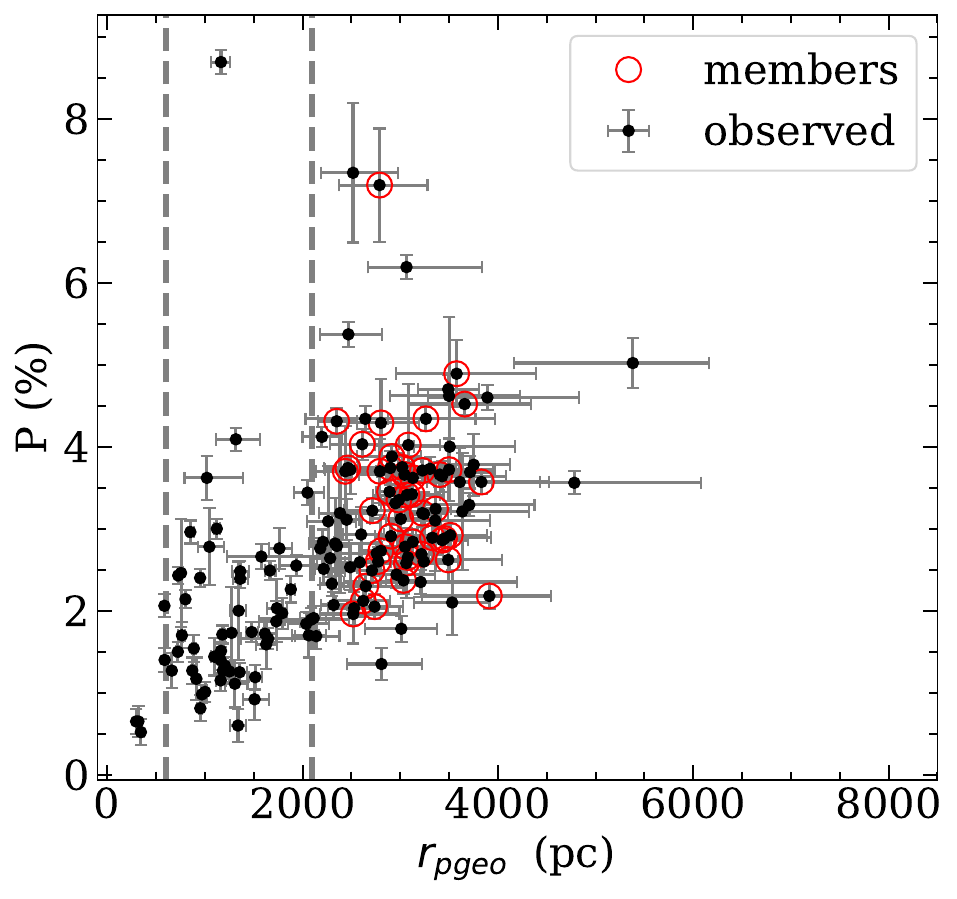}}
\subfigure[]{\label{fig:A2b}\includegraphics[width=0.47\textwidth]{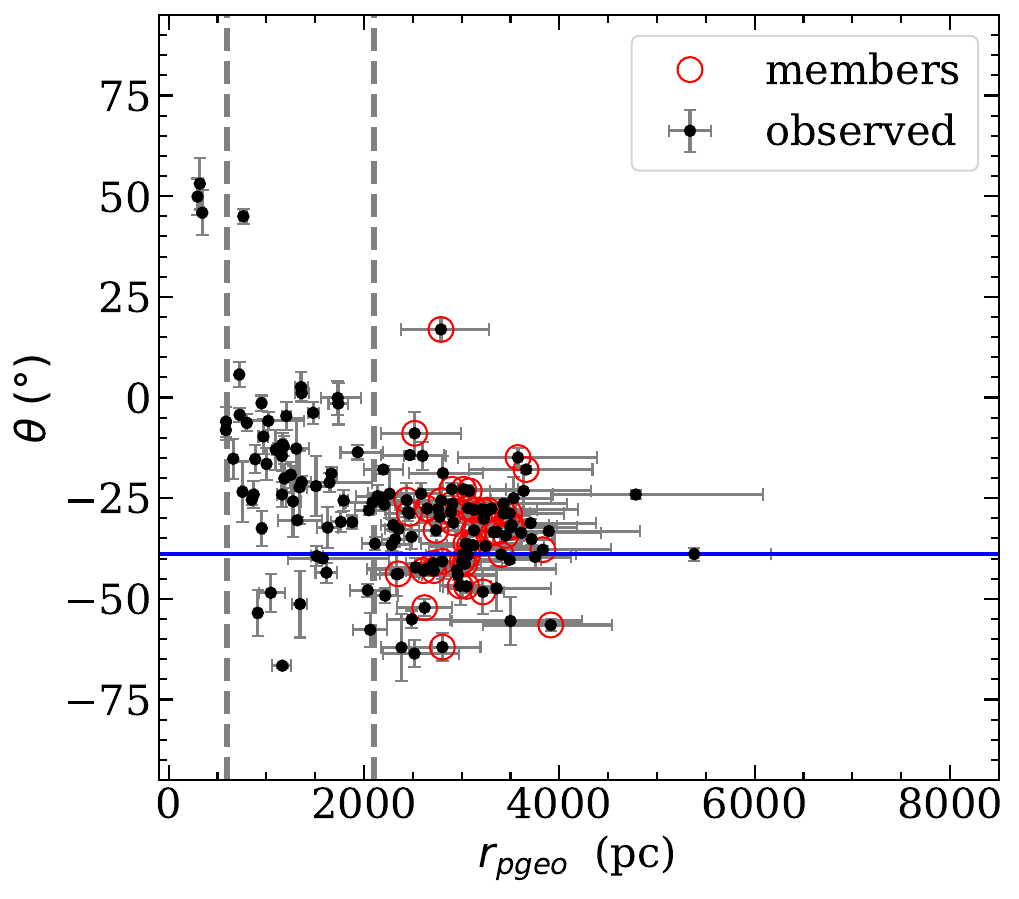}}
 \caption{Same as Fig.~\ref{fig:FigA1} but towards Berkeley~69 cluster direction.}\label{fig:FigA2}
\end{figure*}

\begin{figure*}
\centering
\subfigure[]{\label{fig:A3a}\includegraphics[width=0.44\textwidth]{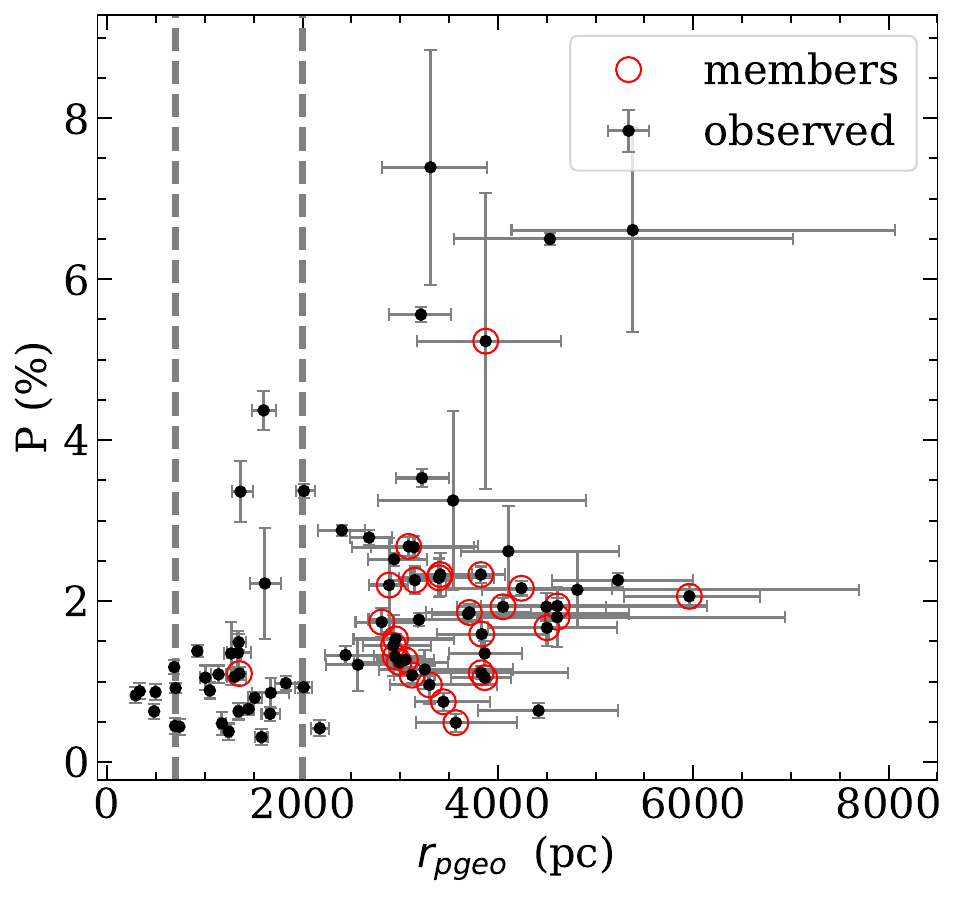}}
\subfigure[]{\label{fig:A3b}\includegraphics[width=0.47\textwidth]{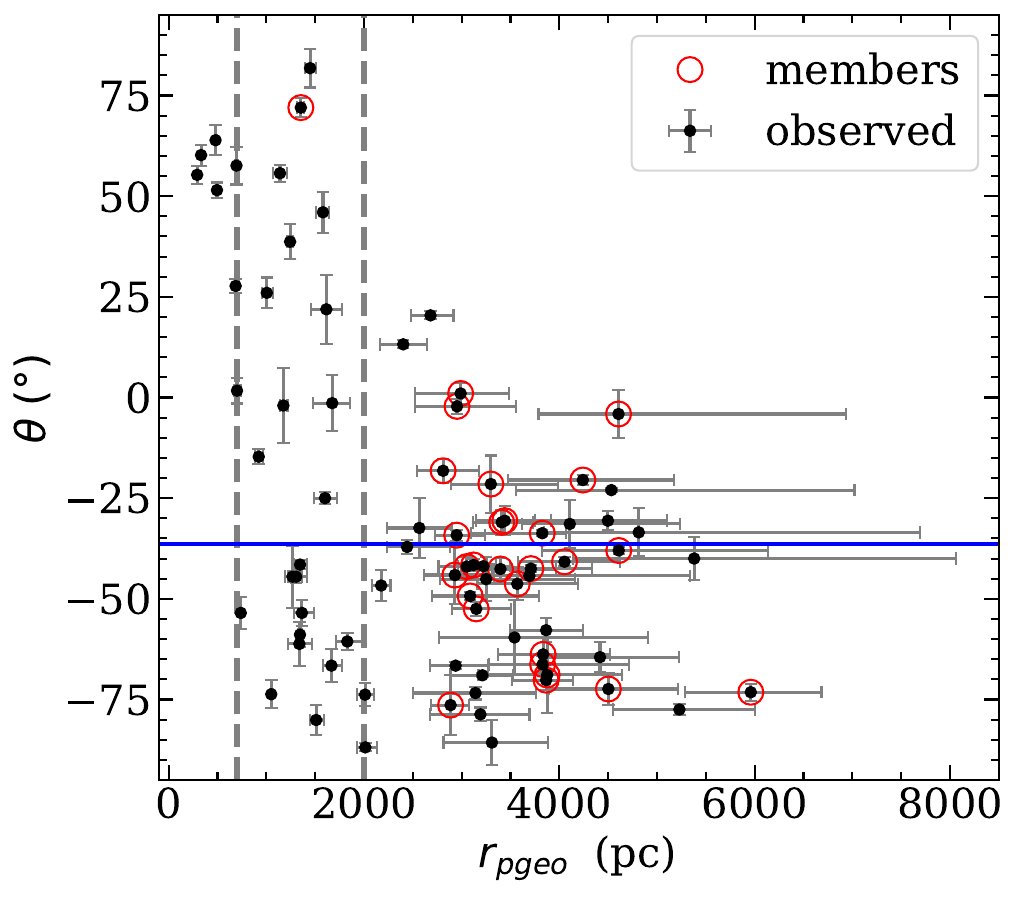}}
 \caption{Same as Fig.~\ref{fig:FigA1} but towards Berkeley~71 cluster direction.}\label{fig:FigA3}
\end{figure*}

\begin{figure*}
\centering
\subfigure[]{\label{fig:A4a}\includegraphics[width=0.44\textwidth]{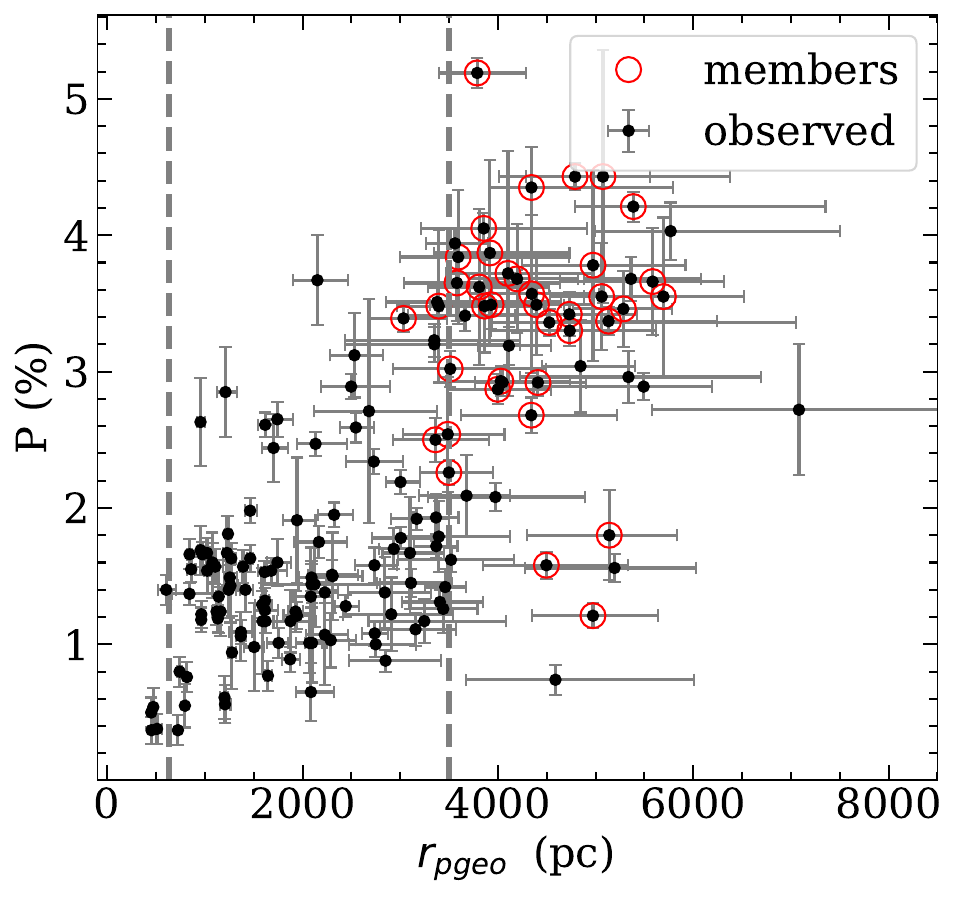}}
\subfigure[]{\label{fig:A4b}\includegraphics[width=0.47\textwidth]{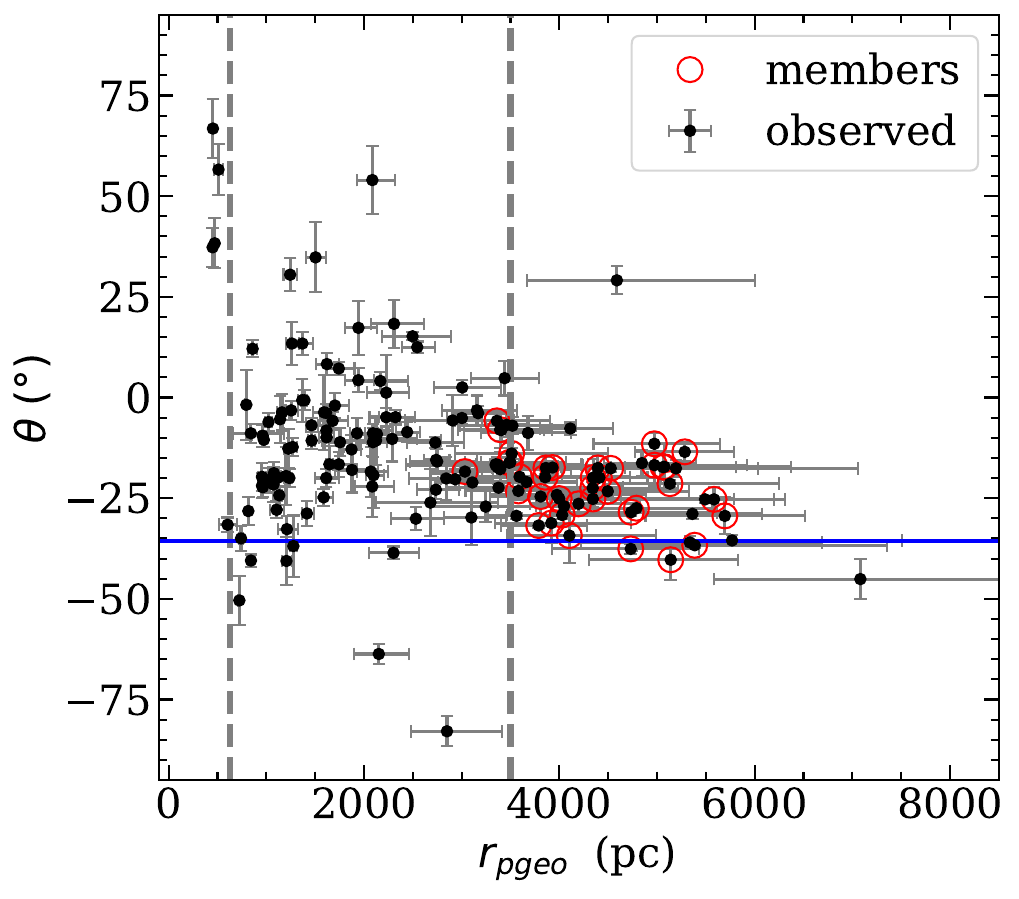}}
 \caption{ Same as Fig.~\ref{fig:FigA1} but towards King~8 cluster direction.}\label{fig:FigA4}
\end{figure*}

\begin{figure*}
\centering
\subfigure[]]{\label{fig:A5a}\includegraphics[width=0.44\textwidth]{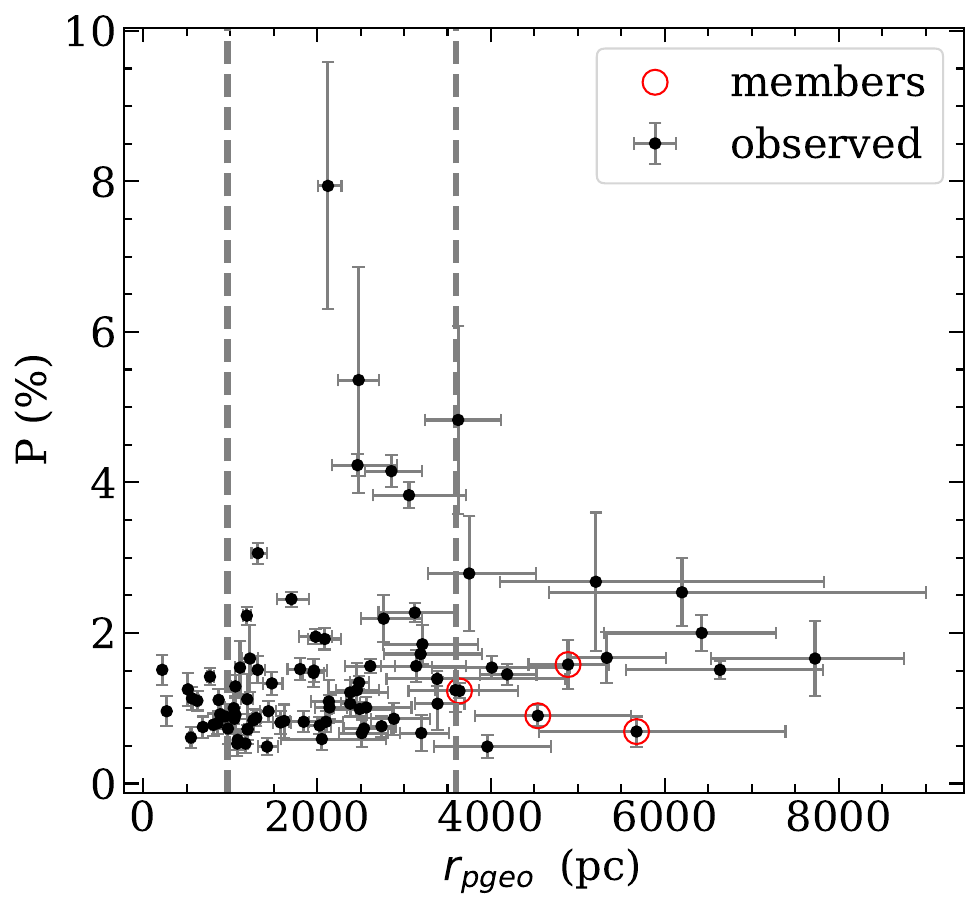}}
\subfigure[]{\label{fig:A5b}\includegraphics[width=0.47\textwidth]{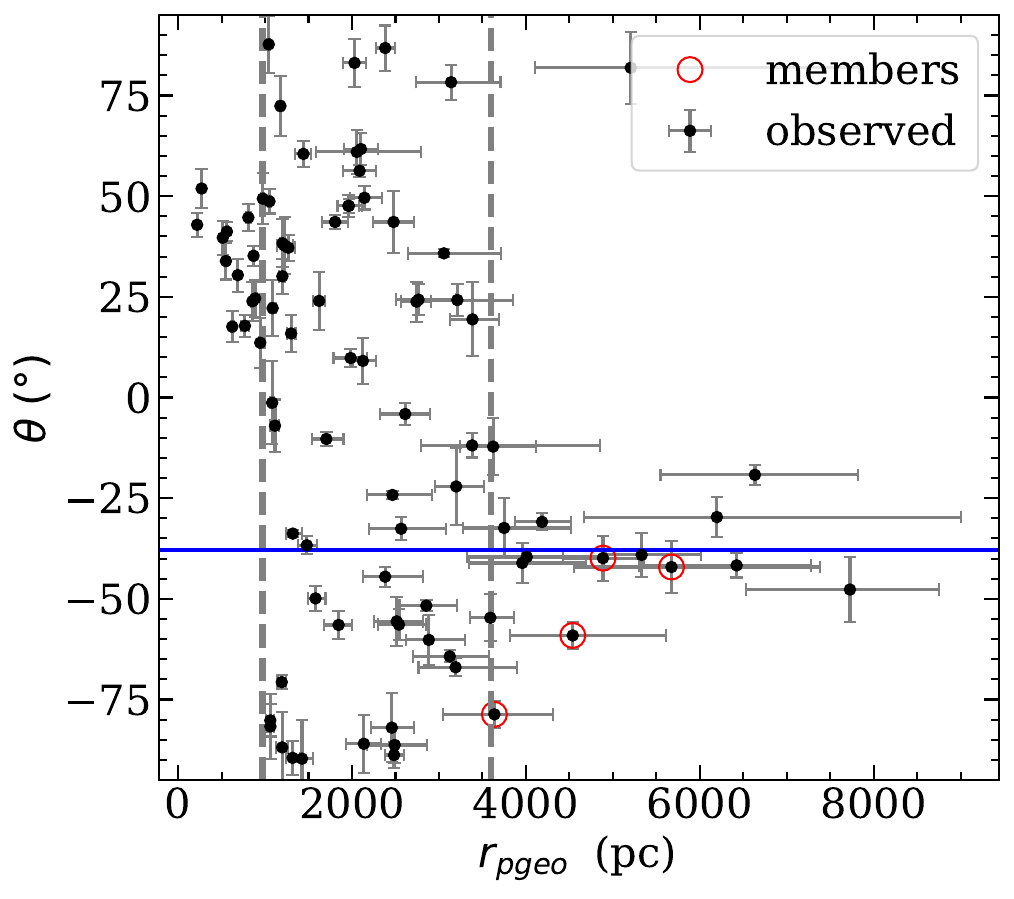}}
 \caption{ Same as Fig.~\ref{fig:FigA1} but towards Berkeley~19 cluster direction.}\label{fig:FigA5}
\end{figure*}

The figures presented herein illustrate the variation of the degree of polarization and polarization angle with distance of stars ($r_{pgeo}$) towards Kronberger~1 (Fig.~\ref{fig:FigA1}), Berkeley~69 (Fig.~\ref{fig:FigA2}), Berkeley~71 (Fig.~\ref{fig:FigA3}), King~8 (Fig.~\ref{fig:FigA4}), and Berkeley~19 (Fig.~\ref{fig:FigA5}). These figures are similar to the top and middle panels of Fig.~\ref{fig:5a}  to Fig.~\ref{fig:5e}, respectively, with the distinction that these incorporate uncertainties in the measured polarization and distance of the stars. The blue solid line in the $\theta$ plot represents the orientation of the Galactic plane towards the corresponding directions. The vertical gray dashed lines in each panel denote the location of the foreground dust layers expected from visual inspection.

\end{appendix}



\end{document}